\documentclass[journal]{IEEEtran}

\usepackage{amsmath,amsfonts}
\usepackage{amsthm}
\usepackage{booktabs}
\usepackage{array}
\usepackage{longtable}
\usepackage{amssymb}
\usepackage{textcomp}
\usepackage{stfloats}
\usepackage{url}
\usepackage{cite}
\usepackage{verbatim}
\usepackage{mathrsfs} 
\usepackage{graphicx}
\usepackage{xcolor}
\usepackage{xurl}
\usepackage{balance}
\usepackage[font=footnotesize]{caption}
\usepackage{subcaption}
\usepackage{algorithm}
\usepackage{algpseudocode} 
\usepackage{balance}
\usepackage{tikz}
\usetikzlibrary{arrows.meta}
\usetikzlibrary{calc}

\newcommand{\real}{\mathbb{R}}

\newcommand{\diag}{\mathrm{diag}}

\newcommand{\sfs}{{\textsf{s}}}

\newcommand{\cF}{\mathcal{F}}
\newcommand{\cN}{\mathcal{N}}
\newcommand{\cG}{\mathcal{G}}

\newcommand{\cL}{\mathcal{L}}

\newcommand{\cP}{\mathcal{P}}

\newcommand{\sfT}{{\textsf{T}}}

\DeclareMathOperator{\range}{range}
\DeclareMathOperator{\Fix}{Fix}
\DeclareMathOperator{\dist}{dist}
\DeclareMathOperator{\prox}{prox}
\DeclareMathOperator*{\argmin}{arg\,min}
\DeclareMathOperator{\proj}{proj}
\newcommand{\e}{\operatorname{e}}

\DeclareSymbolFont{bbold}{U}{bbold}{m}{n}
\DeclareSymbolFontAlphabet{\mathbbold}{bbold}

\DeclareMathAlphabet{\mymathbb}{U}{BOONDOX-ds}{m}{n}
\newcommand{\one}{\mymathbb{1}}
\newcommand{\zero}{\mymathbb{0}}

\newtheorem{lemma}{Lemma}
\newtheorem{proposition}{Proposition}

\newtheorem{definition}{Definition}

\newtheorem{problem}{Problem}
\newtheorem{remark}{Remark}

\allowdisplaybreaks[3]

\title{Dynamic Control Barrier Function  Filters for Frequency Containment in Grid-Following Inverters}

\title{Frequency Containment Filters for Grid-Following Inverters Using Control Barrier Functions}

\title{Decentralized Control Barrier Function  Filters for Frequency Containment With Inverters}

\author{Damola~Ajeyemi and Emiliano~Dall'Anese%
\thanks{D. Ajeyemi is with the Division of Systems Engineering,
Boston University, Boston, MA 02215 USA.  E. Dall'Anese is with the Department of Electrical
and Computer Engineering and the Division of Systems Engineering, Boston University, Boston, MA 02215 USA.
(e-mail: dajeyemi@bu.edu; edallane@bu.edu).}%
\thanks{This work was supported in part by the National Science Foundation award 2546893, and by the U.S. Department of Energy, Office of Electricity, Advanced Grid Modeling program.}}

\begin{document}

\maketitle

\begin{abstract}
This paper addresses the design of safety
filters for grid-following (GFL) inverters in power transmission systems. The goal of the filters is to constrain
frequency nadir and zenith following
generation-demand imbalances, by supplementing primary frequency control.
Using a reduced electromechanical model and control
barrier functions (CBFs), we define an ideal filter
through a convex quadratic program (QP). The ideal QP filter renders the set of admissible frequencies forward invariant; however, its direct implementation is impractical, as it requires centralized implementation and model-based computation. We then use a dual proximal formulation and a two-time-scale implementation to obtain decentralized dynamic filters that can be implemented at each inverter and  
use only local frequency and rate-of-change-of-frequency estimates. To assess the performance and the ability to regulate frequency, we derive  bounds on the distance between state trajectories under the dynamic and ideal CBF filters. The bounds quantify the effects of the filter time constant, initialization, load variation,
and frequency estimation errors,
and provide estimates of possible frequency-limit
violations. We validate the proposed method on the IEEE 39-bus system using a high-fidelity  electromagnetic-transient (EMT) model. 
\end{abstract}

\section{Introduction}

Frequency control in power transmission systems is organized hierarchically, with control
layers operating on different temporal and spatial
scales~\cite{kundur1994power,sauer2017power}.
Following a generation-demand power imbalance, primary control responds
to frequency deviations to arrest the frequency excursion and drive the system to a new equilibrium. Secondary frequency control subsequently acts at the balancing area level to adjust 
power setpoints over tens of seconds to minutes
to restore nominal frequency and scheduled net
power interchanges between areas.
Primary response has traditionally been provided
by turbine-governor droop control of synchronous
generators; however, grid-following (GFL) and
grid-forming (GFM) inverter-based resources (IBRs)
can also provide primary response through
appropriate droop-type controls, subject to available
power headroom and energy
constraints~\cite{dorfler2023control,dallanese2026multitimescale}.

Primary controllers do not guarantee
satisfaction of prescribed transient frequency limits. For IBRs, droop and
synthetic inertia coefficients can be, in principle, engineered
to meet frequency-nadir and steady-state
specifications~\cite{guggilam2018optimizing,zhang2024droop}; transient frequency performance can also be
improved through dynamic droop compensation~\cite{jiang2021dynamic}. However, such parameter-selection and tuning approaches require access
to the IBR controller coefficients and fast communications.

This paper focuses on developing new control mechanisms for GFL inverters to addresses \emph{frequency containment},
defined here as maintaining frequency deviations
 within prescribed bounds following
a generation-demand imbalance.
The approach we propose is to augment primary controllers in GFL inverters with local \emph{filters}, designed based on control barrier
functions (CBFs)~\cite{ames2016control,ames2019control}.
Each CBF filter computes a  correction to the active power reference using local frequency and rate-of-change-of-frequency (RoCoF) estimates, without communication among buses or
centralized computation.

The design starts from an ideal CBF filter defined by a convex quadratic program (QP)~\cite{ames2016control,ames2019control}.
Although the frequency constraints are imposed at individual GFL buses, their derivatives depend
on generator states and power injections through the network; therefore, direct evaluation of the ideal filter requires system-wide
information and joint computation. We derive a local dynamic approximation through
a dual proximal formulation and a two-time-scale dynamic implementation~\cite{dall2026local}, inspired by singular perturbation theory~\cite{kokotovic1976singular}. The implementation is local at each GFL inverter, and complements primary controllers.  We quantify the distance between its plant trajectory and that
of the ideal CBF-controlled system.
These estimates characterize how frequency limit violation  depends on the time-scale parameter of the filter, estimation errors, and time-variability of the loads in the network, quantifying trade-offs between safety  and local implementability.

\emph{Connections to the literature}. 
CBF tools have been explored in power systems to enforce operating limits during transients. For frequency regulation, barrier-based methods have been used to design distributed feedback controllers~\cite{zhang2019distributed} and augment predictive controllers with real-time safety layers, with the objective of limiting frequency deviations while preserving system stability~\cite{zhang2020distributedbilayered}. CBF-based correction mechanisms have also been incorporated into reinforcement learning to address frequency constraints under changing operating conditions and model uncertainty~\cite{wan2025adapsafe2}. In microgrids, distributed barrier certificates and safety filters have been developed to determine admissible power commands and enforce transient operating limits within hierarchical control architectures~\cite{kundu2019distributed,kundu2020transient}. At the converter level, CBFs have been used for
current regulation and current-limit enforcement
in GFL inverters~\cite{pokharel2026gflcbf} and
inverter-based battery storage
systems~\cite{schneeberger2024bessfilter}.
Recent barrier-based control for GFL inverters addresses frequency and current limitation under weak-grid conditions~\cite{michos2026safe}. High-order CBF filters for power system
differential-algebraic models, using an online QP to adjust
governor and exciter references and offline reachability to
certify frequency or voltage constraints for specified
disturbance classes were developed in~\cite{albustami2026power}.

Dynamic local filters based on a two-time-scale
framework were developed in~\cite{dall2026local}
for QPs with a separable  structure and that
admit a unique closed-form optimizer.
In this paper, the power network model yields a coupled
QP whose quadratic cost may be singular, in which
case the solution is not computed in closed form and is non-unique.
The filter design and analysis in~\cite{dall2026local} therefore do
not apply.
We extend the framework through a dual proximal
representation and derive new tracking bounds.

\emph{Contributions}. Our contributions are as follows. 

\noindent \emph{(c1)}  We formulate a CBF filter for frequency containment 
as a convex QP. Depending on the network configuration, the matrix in the quadratic cost of the QP is either positive definite or positive semi-definite and singular. In both cases, we show that the filter renders the set of admissible frequencies forward invariant. 
Through a dual proximal formulation~\cite{vandenberghe2004convex} and a two-time-scale implementation~\cite{kokotovic1976singular}, we derive
a dynamic approximation of this filter that affords a decentralized implementation.
Each GFL inverter implements a scalar update using only local frequency and RoCoF estimates.

\noindent \emph{(c2)} We derive explicit bounds on the distance between
system trajectories under the local dynamic and
ideal filters.
The analysis accounts for initialization,
time-varying inputs, and measurement errors, and
covers both positive-definite and singular cases.
In the singular case, we use regularity of the moving fixed-point set to accommodate non-unique
power corrections of the filter.
The resulting bounds quantify possible frequency-limit
violations. 

\noindent \emph{(c3)} We analyze the effect of limits in the powers that GFL inverters can provide, characterizing the frequency deviations that are due to saturation and limited power headroom. 

\noindent \emph{(c4)} While the filter design is based on a reduced model, we evaluate the performance of the proposed filters on a modified
IEEE 39-bus system using a detailed $dq0$
electromagnetic-transient (EMT) model. To this end, we use the toolbox
\textsc{PowerSimulationsDynamics.jl} (PSID.jl)~\cite{lara2023powersimulationsdynamics}.  The model includes synchronous-machine electromagnetic
dynamics, excitation and turbine-governor controls,
 stabilizers, dynamic transmission lines,
and GFL synchronization, power control, current control,
and LCL filter dynamics.

The remainder of the paper is organized as follows.
Section~\ref{sec:model} presents the problem formulation.
Section~\ref{sec:dynamic_local_filter} develops the CBF filters,
and Section~\ref{sec:filter-analysis} provides their theoretical analysis.
Section~\ref{sec:emt_framework} describes the EMT implementation and numerical experiments.
Section~\ref{sec:conclusion} concludes the paper.

\section{Preliminaries and Problem Formulation}
\label{sec:model}

\emph{Notation:} The set of real numbers is $\real$, and $\real^n_{>0}$
is the set of vectors in $\real^n$ with positive entries. Vector
inequalities hold componentwise, and for $l\leq u$ in $\real^n$,
$[l,u]=\{v\in\real^n:l\leq v\leq u\}$ denotes a box. The symbols
$I_n$ denote the $n\times n$ identity matrix; $\one_n$  and $\zero_n$ denote the
all-ones vector and all-zero vector, respectively; subscripts are omitted when the dimension
is clear. The operator $\operatorname{col}(\cdot)$ stacks its arguments
into a column vector, and $\diag(\cdot)$ forms a diagonal matrix. For a
matrix $Q$, $Q^\dagger$, $\ker Q$, and $\range Q$ denote its
Moore--Penrose inverse, null space, and range. For a symmetric matrix
$Q$, $\lambda_{\min}(Q)$ and $\lambda_{\max}(Q)$ are its extreme
eigenvalues, and $Q\succ0$ ($Q\succeq0$) denotes positive
(semi)definiteness. For $P\succ0$, $\langle v,w\rangle_P=v^\top Pw$
and $\|v\|_P=\langle v,v\rangle_P^{1/2}$; $\|\cdot\|_2$ is the
Euclidean norm and the induced matrix norm. For $z\in\real^m$ and nonempty compact sets
$\mathcal U,\mathcal V\subseteq\real^m$, define the
point-to-set distance $\dist_P(z,\mathcal U)
 :=\inf_{u\in\mathcal U}\|z-u\|_P$ and the Hausdorff distance $d_H^P(\mathcal U,\mathcal V)
 :=\max\bigl\{
   \sup_{u\in\mathcal U}\dist_P(u,\mathcal V),\sup_{v\in\mathcal V}\dist_P(v,\mathcal U)
 \bigr\}$. For a map $T$,
$\Fix(T)=\{z:T(z)=z\}$. Almost everywhere is abbreviated a.e.

\emph{Convex analysis:} Let $C$ be a nonempty closed convex set. Its
support function is $\sigma_C(v)=\sup_{q\in C}v^\top q$, and its
normal cone at $w$ is
$N_C(w)=\{v:v^\top(q-w)\leq0\ \forall q\in C\}$ if $w\in C$, with
$N_C(w)=\varnothing$ otherwise. The Euclidean projection is
$\proj_C(v)=\argmin_{q\in C}\|q-v\|_2$; for scalars $l\leq h$,
$\proj_{[l,h]}(v)=\min\{h,\max\{l,v\}\}$. For a proper closed convex
function $r$, $\partial r$ denotes its subdifferential, and its proximal
map in the metric induced by $P\succ0$
is~\cite{parikh2014proximal}
\begin{equation}
 \prox_r^P(v)=\argmin_q\bigl\{r(q)+\tfrac12\|q-v \|_P^2\bigr\}.
 \label{eq:prelim-prox}
\end{equation}

\emph{Control barrier functions:} Let $\mathcal{D}\subseteq\real^n$ be an open domain containing the safe set, and
let $\mathcal U\subseteq\real^r$ be a nonempty set of
admissible inputs. Consider the control-affine system $\dot x=F(x)+G(x)u$,
where $F:\mathcal{D}\to\real^n$ and $G:\mathcal{D}\to\real^{n\times r}$ are locally
Lipschitz. Let $h:\mathcal{D}\to\real$ be continuously differentiable and
represent the safe set as $\mathcal C_h=\{x\in D:h(x)\geq0\}\subseteq D$. 

\begin{definition}[Control barrier function~\cite{ames2016control}]
\label{def:cbf}
The function $h$ is a control barrier function (CBF) for $\mathcal C_h$
on $D$ if there
exists a locally Lipschitz extended class-$\mathcal K$ function $\alpha$ such that for every $x\in \mathcal{D}$ there is
an input $u\in\mathcal U$ satisfying $\nabla h(x)^\top\bigl(F(x)+G(x)u\bigr)\geq-\alpha\bigl(h(x)\bigr)$.
\end{definition}
Satisfaction of the condition above point-wise in time implies that $\dot x=F(x)+G(x)u$ renders $\mathcal C_h$ forward invariant~\cite{ames2016control}. We also refer the interested reader to higher-order CBFs and adaptive counterparts (see, e.g.,~\cite{xiao2021adaptive}).

\subsection{Problem statement}
\label{subsec:power-system-problem}

We consider a power transmission network with buses $\cN=\{1,\ldots,N\}$.
Synchronous generators (SGs), grid-following (GFL) inverter-based
resources (IBRs), and loads are connected across the network. The sets
$\cG$, $\cF$, and $\cL$ partition $\cN$ into SG buses, buses with a GFL resource, and load-only buses, respectively; loads may
also be connected at buses in $\cG$ and $\cF$. We write
$n_{\cG}=|\cG|$ and $m=|\cF|$, and assume $n_{\cG}\geq1$ and $m\geq1$.
At each bus $i \in \cN$, electrical states of interest on the time scale of primary and secondary control include $\delta_i$, which denotes the phase
angle of the positive-sequence voltage, and the  deviation from the electrical radian synchronous frequency~$\Omega_{\rm s}$ at bus~$i$,  denoted as
\begin{equation}
\label{eq:Omega_def}
\omega_i(t) := \Omega_i(t) - \Omega_{\rm s}. 
\end{equation}
In the context of primary and secondary frequency control,~$\Omega_{\rm s}$ is the nominal frequency  to
which the network is expected to settle with secondary control.

Frequency control is organized
hierarchically~\cite{kundur1994power,sauer2017power,dorfler2023control,dallanese2026multitimescale}. After a
generation-demand imbalance, primary control adjusts active power
injections within seconds to arrest the frequency excursion. We assume that  
frequency-responsive GFL IBRs participate into primary frequency control~\cite{dorfler2023control}. To this end, a GFL IBR (or a collection of IBRs) at bus $i$,
can adjust the nominal active power command as $p_i^{\rm gfl}=\bar p_i^{\rm gfl}-D_{i}\, \omega_i$,
where $D_{i}\geq0$ is the primary response gain, and $\bar p_i^{\rm gfl}$ is the
reference active power set-point accounting for AGC correction  and the dispatch at the tertiary layer.   Primary and secondary control regulate frequency, but their regulation
objectives do not, by themselves, guarantee that local frequency
deviations remain within prescribed limits throughout a transient~\cite{zhang2019distributed,badesa2020optimal}; see also~\cite{dorfler2023control,milano2016frequency}. 

To address this, we propose to augment the GFL active-power
command with a ``power correction'' as:
\begin{equation}
 p_i^{\rm gfl}(t)
 =\bar p_i^{\rm gfl}(t)-D_i \omega_i(t)+z_{i}(t),
 \qquad i\in\cF , 
 \label{eq:supervisory-power-command}
\end{equation}
where the purpose of the power correction $z_{i}$ is to
enforce the prescribed frequency limits. Given this model, the problem addressed in this paper is stated as follows. 

\begin{figure}[t!]
\centering
\definecolor{freqsafe}{RGB}{54,83,145}
\definecolor{freqshade}{RGB}{233,240,246}

\resizebox{.95\columnwidth}{!}{%
\begin{tikzpicture}[
  x=0.95cm,y=0.95cm,font=\footnotesize,
  line cap=round,line join=round,
  axis/.style={draw=black!85,line width=0.65pt,-{Stealth[length=4pt]}},
  bound/.style={draw=black!45,line width=0.55pt,dash pattern=on 3pt off 2.5pt},
  nominal/.style={draw=black!60,line width=1.05pt,dash pattern=on 4pt off 2.5pt},
  filtered/.style={draw=freqsafe,line width=1.25pt}
]
% Frequency limits, admissible band, and nominal-frequency reference.
\fill[fill=freqshade,draw=none] (0,-1.28) rectangle (6.65,0.65);
\draw[bound] (0,0.65) -- (6.65,0.65);
\draw[bound] (0,-1.28) -- (6.65,-1.28);
\draw[black!25,line width=0.45pt] (0,0) -- (6.65,0);
\node[anchor=east] at (-0.10,0.65) {$\overline\omega$};
\node[anchor=east] at (-0.10,0) {$0$};
\node[anchor=east] at (-0.10,-1.28) {$\underline\omega$};
\node[text=black!55] at (4.30,0.32) {Admissible deviations};

% Axes and disturbance time.
\draw[axis] (0,-2.35) -- (0,1.13)
  node[above left,inner sep=2pt] {$\omega_i$};
\draw[axis] (0,-2.35) -- (6.95,-2.35)
  node[right,inner sep=2pt] {$t$};
\draw[black!45,densely dotted,line width=0.6pt]
  (0.65,-2.35) -- (0.65,0.16);
\node[below,inner sep=3pt] at (0.65,-2.35) {$t_0$};

% Nominal response.
\draw[nominal]
  (0,0) -- (0.65,0)
  .. controls (0.80,-0.64) and (1.10,-2.02) .. (1.68,-2.02)
  .. controls (2.14,-2.02) and (2.20,-0.25) .. (2.85,-0.25)
  .. controls (3.25,-0.25) and (3.32,-1.10) .. (3.95,-1.10)
  .. controls (4.47,-1.10) and (4.48,-0.40) .. (5.02,-0.40)
  .. controls (5.48,-0.40) and (5.55,-0.71) .. (5.92,-0.71)
  .. controls (6.20,-0.71) and (6.35,-0.60) .. (6.65,-0.60);

% Filtered response.
\draw[filtered]
  (0,0) -- (0.65,0)
  .. controls (0.80,-0.64) and (1.00,-1.24) .. (1.42,-1.24)
  .. controls (1.94,-1.18) and (2.08,-0.50) .. (2.65,-0.50)
  .. controls (3.25,-0.50) and (3.70,-0.59) .. (4.45,-0.59)
  .. controls (5.10,-0.59) and (5.85,-0.60) .. (6.65,-0.60);

% Legends.
\draw[nominal] (0.85,1.43) -- (1.48,1.43);
\node[anchor=west,inner sep=3pt] at (1.48,1.43) {Primary control};
\draw[filtered] (3.85,1.43) -- (4.48,1.43);
\node[anchor=west,inner sep=3pt] at (4.48,1.43) {With frequency filter};
\end{tikzpicture}%
}
\caption{Idealized illustration of the goal of Problem~\ref{prob:local-frequency-safety}.}
\label{fig:F_concept}
\end{figure}
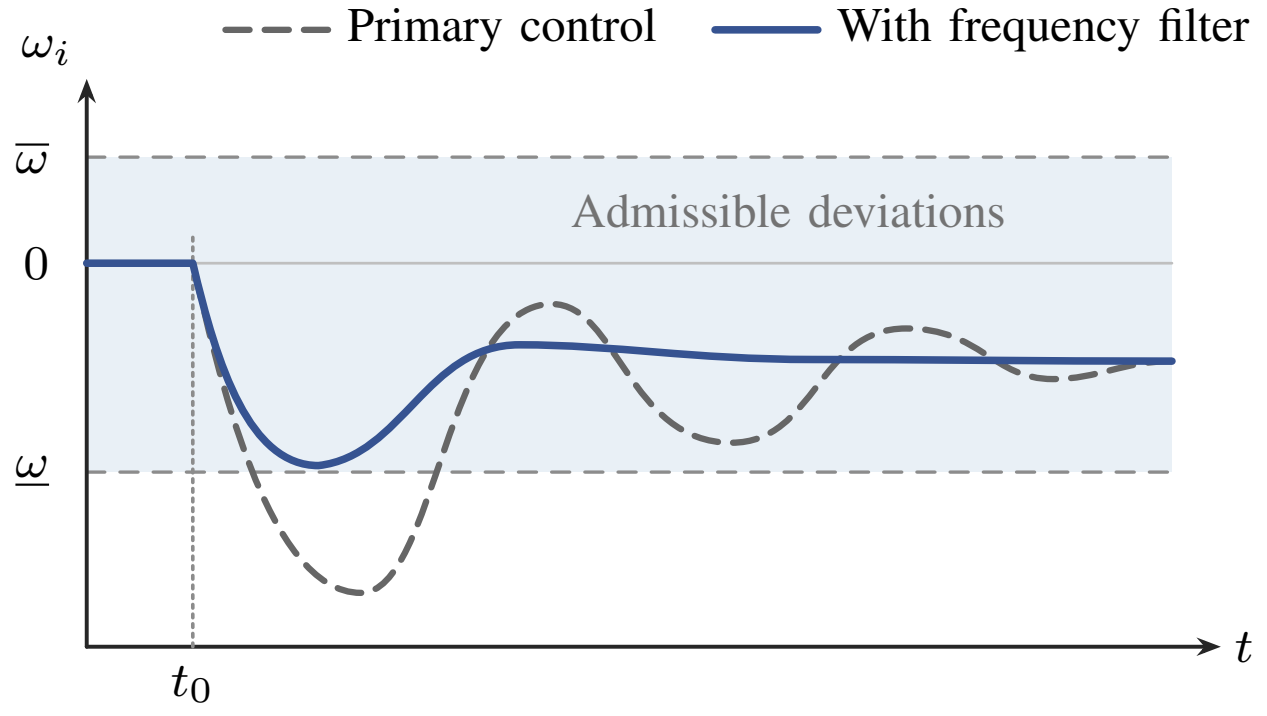

\begin{problem}[Transient frequency containment]
\label{prob:local-frequency-safety}
Given prescribed limits
$\underline\omega<0$ and $\overline\omega > 0$ for the frequency deviation, design a feedback controller $z_{i}(\omega_i)$ at each GFL IBR as in
\eqref{eq:supervisory-power-command}, with the objective of maintaining
\begin{equation}
\underline\omega\leq\omega_i(t)\leq\overline\omega, \,\,\, i\in\cF,
 \label{eq:problem-frequency-limits}
\end{equation}
for all $t \geq 0$.  The feedback controllers shall rely only on local measurements of $\omega_i$. \hfill $\Box$
\end{problem}

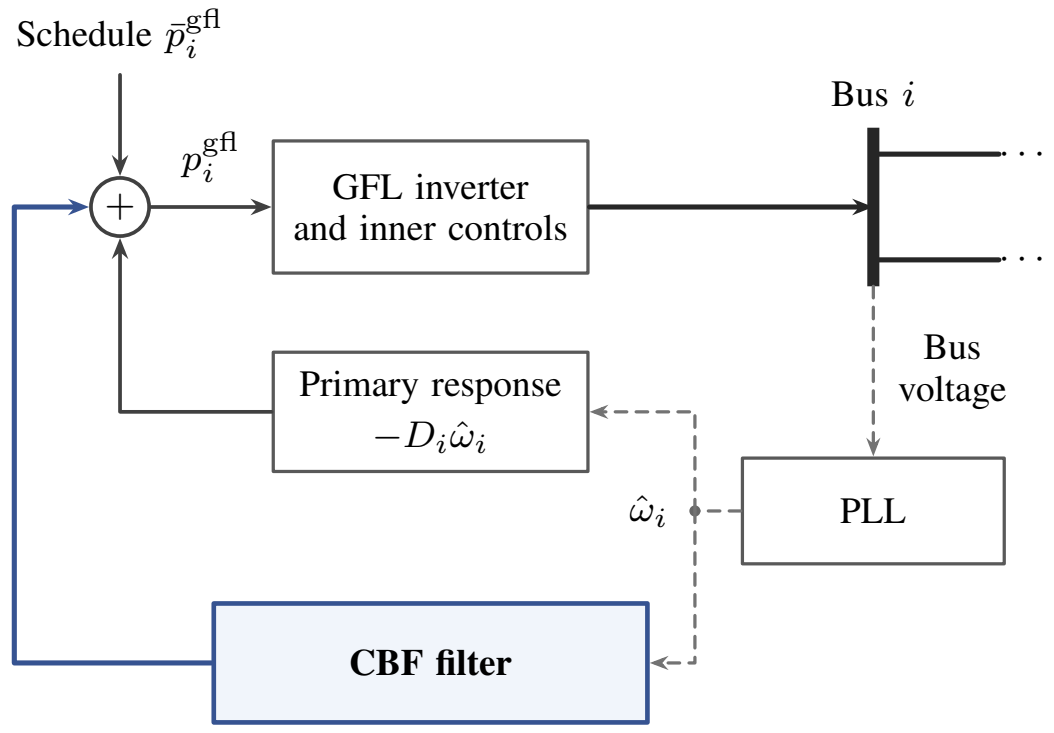
\begin{figure}[t!]
\centering
\begingroup
\definecolor{cbfaccent}{RGB}{54,83,145}
\definecolor{cbfshade}{RGB}{241,245,251}
\resizebox{.785\columnwidth}{!}{%
\begin{tikzpicture}[
  x=1cm,y=1cm,font=\footnotesize,
  line cap=round,line join=round,
  block/.style={draw=black!65,fill=white,line width=0.7pt,
    align=center,inner sep=4pt},
  dynamic/.style={block,draw=cbfaccent,fill=cbfshade,line width=1pt},
  power/.style={draw=black!85,line width=1.1pt,
    -{Stealth[length=4.5pt,width=3.5pt]}},
  electrical/.style={draw=black!85,line width=1.1pt},
  command/.style={draw=black!75,line width=0.8pt,
    -{Stealth[length=4.5pt,width=3.5pt]}},
  correction/.style={draw=cbfaccent,line width=1.1pt,
    -{Stealth[length=4.5pt,width=3.5pt]}},
  measurement/.style={draw=black!55,line width=0.7pt,
    dash pattern=on 2.5pt off 2pt,
    -{Stealth[length=4pt,width=3pt]}},
  signal/.style={fill=white,inner sep=2pt,align=center},
  sum/.style={circle,draw=black!75,fill=white,line width=0.8pt,
    minimum size=0.44cm,inner sep=0pt}
]
% Main actuation path.
\node[sum] (add) at (0.8,3.45) {$+$};
\node[block,text width=2.10cm,minimum height=1.0cm]
  (gfl) at (3.15,3.45) {GFL inverter\\and inner controls};
% Busbar and two outgoing transmission lines.
\coordinate (bus) at (6.50,3.45);
\coordinate (busbottom) at (6.50,2.85);
\draw[draw=black!85,line width=2.6pt,line cap=butt]
  (busbottom) -- (6.50,4.05);
\node[anchor=south,inner sep=2pt] at (6.50,4.13) {Bus $i$};
\draw[electrical] (6.50,3.85) -- (7.45,3.85);
\draw[electrical] (6.50,3.05) -- (7.45,3.05);
\node[inner sep=0pt] at (7.65,3.85) {$\cdots$};
\node[inner sep=0pt] at (7.65,3.05) {$\cdots$};
\draw[command] (add.east)
  -- node[signal,above=3pt] {$p_i^{\rm gfl}$} (gfl.west);
\draw[power] (gfl.east)
  -- node[signal,above=3pt] {} (bus);
% The scheduled reference includes secondary-control adjustments.
\node[align=center,anchor=south] at (0.8,4.45)
  {Schedule $\bar p_i^{\rm gfl}$};
\draw[command] (0.8,4.45) -- (add.north);
% Nominal and supplementary feedback branches.
\node[block,text width=2.10cm,minimum height=0.85cm]
  (droop) at (3.15,1.90)
  {Primary response\\[2pt]$-D_i\hat\omega_i$};
\node[dynamic,text width=3.0cm,minimum height=.90cm]
  (cbf) at (3.15,0)
  {\textbf{CBF filter}};
\draw[command] (droop.west) -- (0.8,1.90) -- (add.south);
\draw[correction] (cbf.west)
  -- (0,0) -- (0,3.45) -- (add.west);
% The PLL measures the voltage at the local bus.
\node[block,text width=1.70cm,minimum height=0.80cm]
  (pll) at (6.50,1.15) {PLL};
\draw[measurement] (busbottom)
  -- node[signal,right=3pt] {Bus\\voltage} (pll.north);
\coordinate (tap) at (5.15,1.15);
\draw[draw=black!55,line width=0.7pt,dash pattern=on 2.5pt off 2pt]
  (pll.west) -- (tap);
\fill[black!60] (tap) circle (1.2pt);
\node[signal,anchor=east] at (5.03,1.15) {$\hat\omega_i$};
\draw[measurement] (tap) -- (5.15,1.90) -- (droop.east);
\draw[measurement] (tap) -- (5.15,0) -- (cbf.east);
\end{tikzpicture}%
}
\endgroup
\caption{Local control architecture at GFL bus $i$.}
\label{fig:F_architecture}
\end{figure}

The inputs $z := \operatorname{col}(z_i)_{i \in \mathcal F}$ adjust the GFL inverters'
active-power commands with the objective of maintaining the local
frequency within the prescribed limits. Figure~\ref{fig:F_concept} provides an idealized illustration of the goal of Problem~\ref{prob:local-frequency-safety} for the case of a loss of generation, while  Figure~\ref{fig:F_architecture} illustrates the desired
architecture associated with
Problem~\ref{prob:local-frequency-safety}. Since the design approach that we pursue is based on tools from the theory of CBF, we will refer to the controller generating $z_i$ to as CBF filter. Requiring the CBF filter to use $\omega_i$ as its only measured
input preserves the local control
structure of conventional primary frequency controllers.  
  
In the next section, we outline the power system model that is utilized for the design of the proposed filters.

\subsection{Power Transmission System Model}
\label{subsec:system_model}

A detailed dynamic model of a power system comprises the generator dynamics
(rotor, field- and damper-winding, excitation, and turbine--governor
dynamics), device-level IBR models (DC link, AC filter, modulation,
inner current loops, phase-locked loop (PLL), and outer power and
voltage controls), load dynamics, ancillary control equipment, and
network electromagnetic
dynamics~\cite{kundur1994power,sauer2017power,yazdani2010voltage}.
Interconnecting these components yields a large-scale nonlinear system of
differential-algebraic equations (DAEs). 

As discussed in, e.g.,\cite{dorfler2023control,dallanese2026multitimescale}, it is useful to separate design models from validation models. In this paper, we will use: 

\begin{itemize}
\item  \emph{High-fidelity dq EMT power transmission system models} for validation; these include dynamics for synchronous generators and their associated control systems, device-level models for IBRs, ancillary control equipment such as voltage compensators, and network dynamics~\cite{lara2023revisiting}. 

\item \emph{A simplified model}  for primary
frequency control dynamics on a time scale of seconds for the design of the proposed CBF filters~\cite{dallanese2026multitimescale}.  
\end{itemize}

In the following, we provide a brief description of the network, generator, and inverter models used for \emph{design} of the CBF filters. For an extensive explanation and a complete derivation of these models, see~\cite{dorfler2023control,ajeyemi2026reachability,dallanese2026multitimescale}. The reader that is familiar with model approximation and reduction techniques  can refer to the notation in Table~\ref{tab:design-model-notation} and jump directly to the  model~\eqref{eq:model-closed-plant-compact}.

\begin{table}[t]
\caption{Notation used for the reduced power system model.}
\label{tab:design-model-notation}
\centering
\footnotesize
\renewcommand{\arraystretch}{1.12}
\setlength{\tabcolsep}{3pt}
\begin{tabular}{@{}p{0.30\columnwidth}p{0.66\columnwidth}@{}}
\toprule
\textbf{Symbol} & \textbf{Meaning} \\
\midrule
$\cN,\cG,\cF,\cL$
& Sets of buses, SG buses, GFL buses, and load buses. \\

$N,n_{\cG},m$
& Numbers of buses, SG buses, and GFL buses. \\

$\delta_i(t),\theta_i(t)$
& Angle and angle relative to reference bus $N$. \\

$\omega_i(t),\hat\omega_i(t)$
& Frequency deviation from $\Omega_{\rm s}$ and its local estimate. \\

$p_{m,i}(t)$ 
& Generator mechanical power. \\

$p_{\ell,i}(t)$, 
& Non-controllable  load at bus $i$. \\

$\bar p_i^{\rm sg}(t),\bar p_i^{\rm gfl}(t)$
& Scheduled SG and GFL active-power references. \\

$M_i,D_i,\tau_i,R_i$
& SG inertia, damping, turbine-governor time constant, and droop coefficient. \\

$D_i$, $i\in\cF$
& GFL primary frequency-response gain. \\

$x(t),d(t)$
& Reduced state and vector of references and loads. \\

$z^\star(x,d)$
& Power correction from centralized CBF filter. \\

$z_i(t)$
& Active-power correction from  dynamic CBF filter at the $i$th GFL inverter \\

$\mathcal B,\mathcal B_{\rm r}$
& Network Laplacian and Kron-reduced Laplacian. \\

$\Phi$
& Frequency-divider matrix~\eqref{eq:frequency-divider-matrix}.  \\

\bottomrule
\end{tabular}

\end{table}

\emph{Network model}. For simplicity of exposition, order the buses as $(\cF,\cL,\cG)$; we also suppress the time index $t$ for notational simplicity. Let $\delta_i$ denote the voltage phase angle at bus $i$ in a reference
frame rotating at the nominal synchronous electrical radian frequency
$\Omega_{\rm s}$. The frequency
deviations are collected in
$\omega:=[\,\omega_1,\ldots,\omega_N\,]^\top\in\real^N$. Without loss of generality, bus $N\in\mathcal G$ is selected as the angle
reference. Define $\theta_i:=\delta_i-\delta_N$, $i=1,\ldots,N-1$, and
$\theta:=[\,\theta_1,\ldots,\theta_{N-1}\,]^\top\in\real^{N-1}$.  As a network approximation,  we consider the DC approximation, with $\mathcal B$ denoting the susceptance matrix of the system (aka  weighted network
Laplacian~\cite{dorfler2013}), and denote as $p^{\rm net}(\theta)$ the vector of net injected powers at the buses. 
 
Hereafter, we assume that the rows and columns of the network Laplacian are grouped 
according to $(\mathcal F, \mathcal L,\cG,)$. Finally, denote as $p_{\ell,i}$ the active-power demand at bus~$i$. At buses with only loads, active-power balance gives $0
=
-p_{\ell,i}
-p_i^{\rm net}(\theta)$.

\emph{Conventional generation model}. Each synchronous generator is represented by the classical electromechanical model with
first-order turbine-governor dynamics
\cite{dallanese2026multitimescale}. For
$i\in\mathcal G\setminus\{N\}$, the relative rotor-angle dynamics satisfy
$\dot\theta_i=\omega_i-\omega_N$, and the frequency and mechanical power
dynamics are 
\begingroup
\setlength{\abovedisplayskip}{3pt}
\setlength{\belowdisplayskip}{3pt}
\begin{subequations}
\label{eq:sg_reduced_model}
\begin{align}
M_i\dot\omega_i
&=
-D_i\omega_i+p_{m,i}
-p_{\ell,i}-p_i^{\rm net}(\theta),
\\
\tau_i\dot p_{m,i}
&=
-p_{m,i}+ \bar{p}_i^{\rm sg}-R_i^{-1}\omega_i, 
\end{align}
\end{subequations}
\endgroup
where $M_i > 0$, $D_i > 0$, $\tau_i > 0$, and $R_i$ denote the inertia, damping,
turbine-governor time constant, and frequency-power droop coefficient,
respectively; $p_{m,i}$ is the turbine mechanical power, and
$\bar{p}_i^{\rm sg}$ is the reference (supplied by
secondary control and economic dispatch~\cite{kundur1994power}).

\emph{GFL inverters}. Assume that the GFL inverters remain synchronized and that their
PLL and active-power tracking dynamics are stable and sufficiently
fast relative to the dynamics retained for filter design~\cite{dorfler2023control}. Neglecting
these tracking transients, the injected active power follows
\eqref{eq:supervisory-power-command}. Since the reduced GFL model does not have an independent
frequency state, an additional relation is needed to specify
the local frequency driving its frequency  response. To this end, we use the so-called frequency-divider
approximation~\cite{milano2016frequency}:
\begin{equation}
 \omega_{\cF} \approx \Phi\omega_{\cG},
 \label{eq:model-frequency-proxy}
\end{equation}
where $\omega_{\cG}:=\operatorname{col}(\omega_i)_{i\in\cG}$,
with the angle-reference bus $N$ listed last, and $\omega_{\cF}:=\operatorname{col}(\omega_i)_{i\in\cF}$. To show the expression of $\Phi\in\real^{m\times n_{\cG}}$, let
$\mathcal A:=\cF\cup\cL$. With the ordering considered in the paper, the network Laplacian can be written with  the block form $\mathcal B=
 \begin{bmatrix}
  \mathcal B_{\mathcal A\mathcal A}
    & \mathcal B_{\mathcal A\cG}\\
  \mathcal B_{\cG\mathcal A}
    & \mathcal B_{\cG\cG}
 \end{bmatrix}$,
where $\mathcal B_{\mathcal X\mathcal Y}$ denotes the
submatrix with rows indexed by $\mathcal X$ and columns
indexed by $\mathcal Y$. The frequency-divider matrix $\Phi$ is then given by~\cite{milano2016frequency,dallanese2026multitimescale}
\begin{equation}
 \Phi:=
 -S_{\cF}
   \mathcal B_{\mathcal A\mathcal A}^{-1}
   \mathcal B_{\mathcal A\cG},
 \label{eq:frequency-divider-matrix}
\end{equation}
where  $S_{\cF}:=[\,I_m\ \zero_{m\times|\cL|}\,]$.
We note that for a connected network and 
$n_{\cG}\geq1$, $\mathcal B_{\mathcal A\mathcal A}$ is positive
definite~\cite{dorfler2013}. Moreover,  $\Phi$ is
entrywise nonnegative and satisfies
$\Phi\one_{n_{\cG}}=\one_m$~\cite{milano2016frequency}.

\emph{Reduced model}. Hereafter, for angles, frequencies, and powers, a bus-set subscript
denotes stacking over the corresponding buses; for example $p_{m,\cG} := \operatorname{col}(p_{m,i})_{i\in\cG}$. One exception is the vector of relative angles $\theta_{\cG}:=
 \operatorname{col}(\theta_i)_{i\in\cG\setminus\{N\}}$, where the angle-reference bus is obviously not considered.
Next, define what will be our dynamic state vector in the reduced model: 
\[
 x:=\operatorname{col}
       (\theta_{\cG},\omega_{\cG},p_{m,\cG})
 \in\real^{3n_{\cG}-1}.
\]
Using~\eqref{eq:model-frequency-proxy}  and assuming that $\theta_N=0$ without loss of generality, yields the linear time-invariant model 
\cite{dallanese2026multitimescale}:
\begin{subequations}
\label{eq:model-closed-plant}
\begin{align}
 \dot\theta_{i}
 &= \omega_{i} - \omega_N, \,\,\, i \in \cG \setminus\{N\} , 
 \label{eq:model-relative-angles}\\
 M_{\cG}\dot\omega_{\cG}
 ={}&-\mathcal B_{\rm r}
       \begin{bmatrix}\theta_{\cG}\\0\end{bmatrix}
       +p_{m,\cG}-D_{\rm e}\omega_{\cG}
 \nonumber\\
 &-p_{\ell,\cG}
   +(\mathcal B_{\mathcal A\mathcal A}^{-1}
       \mathcal B_{\mathcal A\cG})^\top p_{\ell,\mathcal A} +\Phi^\top(\bar p_{\cF}^{\rm gfl}+z),
 \label{eq:model-reduced-swing}\\
 T_{\cG}\dot p_{m,\cG}
 &=-p_{m,\cG}-R_{\cG}^{-1}\omega_{\cG}
       +\bar p_{\cG}^{\rm sg},
 \label{eq:model-governor}
\end{align}
\end{subequations}
where $D_{\rm e}:=D_{\cG}+\Phi^\top D_{\cF}\Phi$ and
\begin{equation}
 \mathcal B_{\rm r}:=
 \mathcal B_{\cG\cG}
 -\mathcal B_{\cG\mathcal A}
  \mathcal B_{\mathcal A\mathcal A}^{-1}
  \mathcal B_{\mathcal A\cG}
 \in\real^{n_{\cG}\times n_{\cG}}
 \label{eq:model-kron-reduction}
\end{equation}
is the Kron-reduced network Laplacian~\cite{dorfler2013}; the parameter matrices are $M_{\cG}:=\diag(M_i)_{i\in\cG}$,
$T_{\cG}:=\diag(\tau_i)_{i\in\cG}$,
$R_{\cG}:=\diag(R_i)_{i\in\cG}$,
$D_{\cG}:=\diag(D_i)_{i\in\cG}$, and
$D_{\cF}:=\diag(D_i)_{i\in\cF}$. For convenience, the notation is summarized in
Table~\ref{tab:design-model-notation}.

The nominal GFL response contributes the damping
$\Phi^\top D_{\cF}\Phi\succeq0$, while the supplementary
corrections enter through $\Phi^\top z$.
Collecting the scheduled powers and loads in
\[
 d:=\operatorname{col}
       (\bar p_{\cG}^{\rm sg},\bar p_{\cF}^{\rm gfl},p_{\ell,\cN}),
\]
and with standard manipulations, the reduced model~\eqref{eq:model-closed-plant} can be compactly written as
\begin{equation}
 \dot x=Ax+Bd+G z, \,\,\,\,\,\,\,\,  G:=
 \begin{bmatrix}
  \zero_{(n_{\cG}-1)\times m}\\
  M_{\cG}^{-1}\Phi^\top\\
  \zero_{n_{\cG}\times m}
 \end{bmatrix} \, .
 \label{eq:model-closed-plant-compact}
\end{equation}
We will use the model~\eqref{eq:model-closed-plant-compact} for the design of the CBF filter in the next section. The performance of the filter will then be  validated using high-fidelity EMT models.

\section{Dynamic Filter for Frequency Containment}
\label{sec:dynamic_local_filter}

This section outlines the design of the proposed CBF filter.

\subsection{Frequency Constraints and Ideal CBF Filter}

We start by formalizing the frequency constraints.  Let $C:=
 [\,\zero_{m\times(n_{\cG}-1)}\ \Phi\
    \zero_{m\times n_{\cG}}\,]$, so that we have $\omega_{\cF}=C x$. For each $i\in\cF$, let
$c_i^\top$ denote the corresponding row of $C$ and define
\begin{subequations}
\begin{align}
 \underline{h}_i(x) & :=c_i^\top x-\underline\omega, \\
 \overline{h}_i(x)& :=\overline\omega-c_i^\top x.
\end{align}
\end{subequations}
The constraints on the frequency deviation  are equivalent to
$\underline{h}_i(x)\geq0$ and $\overline{h}_i(x)\geq0$.
Along trajectories of~\eqref{eq:model-closed-plant-compact},
\begin{subequations}
 \label{eq:frequency-barrier-derivatives}
 \begin{align}
 \dot{\underline{h}}
 &=C(Ax+Bd)+\Gamma z,\\
 \dot{\overline{h}}
 &=- C(Ax+Bd) - \Gamma z,
 \end{align}
\end{subequations}
where 
\begin{equation}
 \Gamma:=CG
 =\Phi M_{\cG}^{-1}\Phi^\top.
 \label{eq:Gamma}
\end{equation}
These derivatives yield affine CBF inequalities that will be used to construct the CBF filter. Before proceeding, we provide a result on $\Gamma$ that will be useful later on. 

\begin{proposition}[Properties of the matrix $\Gamma$]
\label{prop:gamma}
The matrix $\Gamma = \Phi M_{\cG}^{-1}\Phi^\top$ is symmetric
positive semidefinite, with $\ker(\Gamma)=\ker(\Phi^\top)$ and  $\operatorname{rank}(\Gamma)=\operatorname{rank}(\Phi)$. Consequently, $\Gamma\succ0$ if and only if $\Phi$ has full
row rank; in particular, $\Gamma$ is singular whenever
$m>n_{\cG}$. \hfill $\Box$
\end{proposition}

Since the inertia coefficients are positive,
$M_{\cG}\succ0$; moreover, $\Phi$ is non-negative~\cite{milano2016frequency}. The proof amounts to showing that, for every $v\in\real^m$, $v^\top\Gamma v
 =\bigl\|M_{\cG}^{-1/2}\Phi^\top v\bigr\|_2^2\geq0$,
with equality if and only if $\Phi^\top v=0$. While the design of the filter will not depend on whether $\Gamma$ is positive semidefinite or positive definite, these two cases will require two dedicated approaches for the analysis.   

\begin{figure*}[!t]
\centering
\input{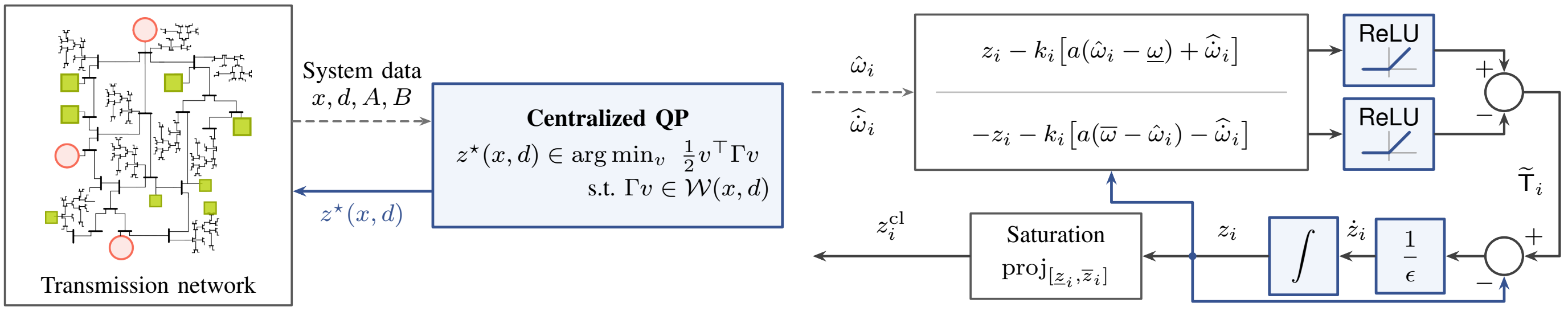}
\caption{Ideal centralized and local dynamic CBF filters. (a)
The centralized QP uses system-wide data $(x,d)$ to compute the
vector of corrective powers. (b) The dynamic filter at bus $i$ uses
only local frequency and RoCoF estimates. Section~\ref{sec:containment} considers the case where the saturation is not active, and $z_i$ enters the inverter as the input. Section~\ref{sec:input_constraints} then considers the effect of power limits.}
\label{fig:local-filter-implementation}
\end{figure*}

The set associated with the frequency limits is
\begin{equation}
 \mathcal C:=
 \left\{x\in\real^{3n_{\cG}-1}:
 \underline\omega\one_m\leq C x
 \leq\overline\omega\one_m\right\}.
 \label{eq:frequency-safe-set}
\end{equation}
Using Definition~\ref{def:cbf} for 
$\underline{h}_i$ and $\overline{h}_i$ with a linear function $\alpha(h) = a h$, $a>0$, the conditions
$\dot{\underline{h}}_i\geq-a \underline{h}_i$ and
$\dot{\overline{h}}_i\geq-a \overline{h}_i$ are equivalent to
\begin{equation}
 \underline\beta(x,d)\leq\Gamma z
 \leq\overline\beta(x,d),
 \label{eq:cbf-box}
\end{equation}
where
\begin{align}
 \underline\beta(x,d)
 &:=-a(C x-\underline\omega\one_m)-C(Ax+Bd),
 \nonumber\\
 \overline\beta(x,d)
 &:=a(\overline\omega\one_m-C x)-C(Ax+Bd).
 \label{eq:beta}
\end{align}
Using~\eqref{eq:cbf-box}, we define the ideal CBF filter as a feedback law $z^\star(x,d)$; in particular, for given $x,d$, the CBF filter selects an
active-power correction $z^\star(x,d)$ from the set of minimizers of the following convex quadratic program (QP):
\begin{equation}
 \begin{aligned}
z^\star(x,d) \in \arg \min_{v\in\real^m}
 &\quad \tfrac12 v^\top\Gamma v\\
 \operatorname{s.\ to}
 &\quad \underline\beta(x,d)\leq\Gamma v             \leq\overline\beta(x,d) .
 \end{aligned}
 \label{eq:static-qp}
\end{equation}
Solving the QP~\eqref{eq:static-qp} in the optimization variable $v$ point-wise in time yields the closed-loop system:
\begin{equation}
 \dot x(t)=Ax(t)+Bd(t)+G z^\star(x(t),d(t)) \, ~~ t \geq 0.
 \label{eq:ideal-closed-loop}
\end{equation}
This closed-loop system is illustrated in Fig.~\ref{fig:local-filter-implementation}(a). We note that, when $\Gamma$ is singular,~\eqref{eq:static-qp} may have multiple solutions; however, since $\ker(\Gamma)=\ker (G)$, every optimizer induces the same  input $G z^\star(x,d)$. We also note that the vector $\Phi^\top v$ represents the equivalent
power correction entering the reduced  model~\eqref{eq:model-reduced-swing}; the cost of~\eqref{eq:static-qp} therefore  penalizes the correction through
$v^\top\Gamma v
=\|M_{\cG}^{-1/2}\Phi^\top v\|_2^2$, which measures this correction with inverse-inertia
weights as $v^\top\Gamma v
 =\sum_{j\in\cG}
   \frac{\bigl[(\Phi^\top v)_j\bigr]^2}{M_j}$. Thus, the  filter~\eqref{eq:static-qp}  minimizes the intervention in the
reduced dynamics, measured in this metric, subject to the
frequency constraints. We next establish the frequency-containment guarantee of the
ideal filter~\eqref{eq:static-qp}.

\begin{proposition}[Forward invariance under the ideal filter]
\label{prop:ideal-invariance}
Let $d(t)$ be locally integrable, and let $z^\star(x,d)$
be any optimizer of~\eqref{eq:static-qp}. The closed-loop system~\eqref{eq:ideal-closed-loop} has a unique forward-complete state trajectory for every
initial condition.
Moreover,~\eqref{eq:ideal-closed-loop} renders the set $\mathcal C$  forward invariant. \hfill $\Box$
\end{proposition}

The proof is provided in the Appendix.  The ideal filter~\eqref{eq:static-qp} provides a \emph{theoretical
benchmark}: its implementation is not practical, as its centralized information requirements are
incompatible with the local architecture of conventional
primary frequency control. Evaluating its constraints requires
the full state $x$ and the loads and power references
collected in $d$, while $\Gamma$ couples the corrective power
injections across GFL buses. 
This motivates the dynamic filter developed next, which
uses only local measurements.

We note that, for now, we neglect GFL active-power limits, assuming that each
resource can deliver the required corrective power.
These constraints 
will be considered in  Section~\ref{sec:input_constraints}.

\subsection{Dynamic Local Filter for Frequency Containment}
\label{sec:containment}

Building on the ideal CBF filter~\eqref{eq:static-qp}, we
propose a dynamic CBF filter that uses local frequency measurements. We first present the filter, and then explain how it provides a dynamic  approximation of~\eqref{eq:static-qp}.

\vspace{.1cm}

\emph{Proposed dynamic filter}.
At each bus $i\in\cF$, let $\hat\omega_i$ denote a measurement of the local
frequency deviation, and let
$\widehat{\dot\omega}_i$ be an estimate of the rate-of-change of $\omega_i$
(obtained by processing the same local frequency estimate from the PLL). Introduce an auxiliary dynamic state $z_i$, which will replace the ideal one 
$z^\star_{i}(x,d)$, and let $k_i>0$ are tuning parameters, for $i \in \cF$.  
The proposed dynamic CBF filter, interconnected with the power systems model,
is given by
\begin{subequations}
\label{eq:local-dynamic-filter}
\begin{align}
 \dot x(t)
 &=Ax(t)+Bd(t)+Gz(t),
 \label{eq:local-dynamic-filter-plant}\\
  \epsilon\dot z_i(t) &=-z_i(t)+ \widetilde{\sfT}_i(z_i(t),\widehat{\dot\omega}_i(t),\widehat{\omega}_i(t)) , ~~~~ i\in\cF, 
 \label{eq:local-dynamic-filter-z}
\end{align}
\end{subequations}
where $\epsilon>0$ enforces a time-scale separation as in classical two-time-scale models~\cite{kokotovic1976singular} and $\widetilde{\sfT}_i(z_i,\widehat{\dot\omega}_i,\widehat{\omega}_i)$ is defined as: 
\begin{align}
\widetilde{\sfT}_i(z_i,\widehat{\dot\omega}_i,\widehat{\omega}_i) =  & ~ \textsf{ReLU}\!\left(z_i+k_i\bigl[-a(\hat\omega_i-\underline\omega)
                    -\widehat{\dot\omega}_i\bigr]
 \right) \nonumber \\
 & - \! \textsf{ReLU}\!\left(
   -z_i-k_i\bigl[a(\overline\omega-\hat\omega_i)
                    -\widehat{\dot\omega}_i\bigr]
 \right)
  \label{eq:local-dynamic-filter-z-closedform}
\end{align}
with $\textsf{ReLU}(x) = \max\{0,x\}$. Differently from $z^*(x,d)$,  $\widetilde{\sfT}_i(z_i,\widehat{\dot\omega}_i,\widehat{\omega}_i)$ is implementable at each GFL inverter $i \in \cF$, as it uses only its local state $z_i$ and local
measurements $\widehat{\omega}_i$, without  communication among buses. This architecture is illustrated in Fig.~\ref{fig:local-filter-implementation}(b).

Next, we show how we derived the dynamic filter~\eqref{eq:local-dynamic-filter-z} as a dynamic approximation of the ideal filter~\eqref{eq:static-qp}. First, a remark is in order. 

\vspace{.1cm}

\begin{remark}[Estimation of $\dot \omega_i$]
\label{rem:local_rocof_estimation}
The
dynamic CBF filter~\eqref{eq:local-dynamic-filter-z}
uses estimates of $\omega_i$ and 
$\widehat{\dot{\omega}}_i$. Since direct differentiation of the frequency
signal amplifies high-frequency measurement noise, one may utilize the so-called dirty derivative~\cite{sira2019equivalence}. For each
$i\in\mathcal F$, let $\chi_i$ denote an auxiliary estimator state; then, the dirty derivative method involved the following dynamics
\begin{subequations}
\label{eq:local_rocof_estimator}
\begin{align}
\tau_{d,i}\dot{\chi}_i
&=
-\chi_i+\omega_i,
\label{eq:local_rocof_estimator_state}
\\
\widehat{\dot{\omega}}_i
&=
\frac{\omega_i-\chi_i}{\tau_{d,i}},
\label{eq:local_rocof_estimate}
\end{align}
\end{subequations}
where $\tau_{d,i}>0$ is a time constant and
$\tau_{d,i}^{-1}$ is its cutoff bandwidth.
\hfill$\Box$
\end{remark}

\vspace{.1cm}

\emph{Derivation of~\eqref{eq:local-dynamic-filter-z} from the QP~\eqref{eq:static-qp}}. 
The derivation of~\eqref{eq:local-dynamic-filter-z} follows three steps: \emph{(s1)} we first consider the dual of~\eqref{eq:static-qp} and show that
its optimizers are fixed points of a proximal map;  \emph{(s2)} we consider a dynamic    approximation of the proximal map; and, \emph{(s3)} we replace the system model with measurements. While similar ideas were pursued in~\cite{dall2026local}, the steps \emph{(s1)} and \emph{(s2)} are unique to the approach proposed here. 

First, for fixed $(x,d)$, define the constraint box
$\mathcal W(x,d)
 :=[\underline\beta(x,d),\overline\beta(x,d)]$.
Introduce $w=\Gamma v$ and associate a multiplier
$z\in\real^m$ with the equality $w-\Gamma v=0$ (we will show the link between $z$ and $z^*$ shortly).
The Lagrangian function is
\begin{equation}
 L(v,w,z)
 =\frac12v^\top\Gamma v
   +z^\top(w-\Gamma v),
 \,\, w\in\mathcal W(x,d).
\end{equation}
Minimizing over $v$ and $w$ gives the dual function
$g(z;x,d)=-\frac12 z^\top\Gamma z-r(z;x,d)$, where
\begin{equation}
 \begin{aligned}
 r(z;x,d)
 &:=\max_{w\in\mathcal W(x,d)}(-z^\top w)\\
 &=\sum_{i\in\cF}
   \max\{-\underline\beta_i(x,d)z_i,
          -\overline\beta_i(x,d)z_i\}.
 \end{aligned}
 \label{eq:box-support-function}
\end{equation}
Thus, $r(z;x,d)$ is the support function of
$\mathcal W(x,d)$ evaluated at $-z$.
By Lagrangian duality
\cite[Secs.~5.1--5.2]{vandenberghe2004convex},
the dual problem maximizes $g$, or equivalently,
minimizes its negative:
\begin{equation}
 \min_{z\in\real^m}
 \left\{\frac12z^\top\Gamma z+r(z;x,d)\right\}.
 \label{eq:dual-problem}
\end{equation}
The midpoint of $\mathcal W(x,d)$ belongs to
$\range(\Gamma)$ and lies strictly inside the box.
Consequently, the  QP is strictly feasible.

Collect the filter gains in
$K:=\diag(k_i)_{i\in\cF}\succ0$.
Using the weighted proximal
operator~\eqref{eq:prelim-prox}, define
\begin{equation}
 \sfT(z;x,d)
 :=\prox_{r(\cdot;x,d)}^{K^{-1}}(z-K\Gamma z).
 \label{eq:prox-map}
\end{equation}

Since $r(z;x,d)$ is separable and $K$ is diagonal,
the proximal map can be evaluated componentwise as
\begin{subequations}
\label{eq:scalar-proximal-map}
\begin{align}
 a_i(z;x,d)
 &=z_i+k_i[\underline\beta_i(x,d)-(\Gamma z)_i],\\
 b_i(z;x,d)
 &=z_i+k_i[\overline\beta_i(x,d)-(\Gamma z)_i],\\
 \sfT_i(z;x,d)
 &=\proj_{[a_i(z;x,d),b_i(z;x,d)]}(0),
 \qquad i\in\cF,
\end{align}
\end{subequations}
where $\sfT_i(z;x,d)$ is the $i$th component of $\sfT(z;x,d)$ and $(\Gamma z)_i$ is the $i$th entry of the vector $\Gamma z$.  Note that since $b_i(z;x,d)
 - a_i(z;x,d)
 =k_i a(\overline\omega-\underline\omega)\geq 0$,
the projection interval is always well defined.  Additionally, we note that the projection $\proj_{[a_i(z;x,d),b_i(z;x,d)]}(0)$ can be computed in closed form as 
$\proj_{[a_i,b_i]}(0) = \textsf{ReLU}(a_i) - \textsf{ReLU}(-b_i)$. 
Along trajectories of
\eqref{eq:model-closed-plant-compact},
we have $\dot\omega = C(A x + B d)+ \Gamma z$.
Using~\eqref{eq:beta}, we therefore obtain
\begin{equation}
 \begin{aligned}
 \underline\beta_i(x,d)-(\Gamma z)_i
 &=-a(\omega_i-\underline\omega)-\dot\omega_i,\\
 \overline\beta_i(x,d)-(\Gamma z)_i
 &=a(\overline\omega-\omega_i)-\dot\omega_i,
 \end{aligned}
 \label{eq:local-cbf-residuals}
\end{equation}
for every $i\in\cF$.

For brevity, let  $\widetilde a_i(z_i,\widehat{\dot\omega}_i,\widehat{\omega}_i)
 =z_i+k_i\bigl[-a(\hat\omega_i-\underline\omega)
                    -\widehat{\dot\omega}_i\bigr]$, and 
 $\widetilde b_i(z_i,\widehat{\dot\omega}_i,\widehat{\omega}_i)
 =z_i+k_i\bigl[a(\overline\omega-\hat\omega_i)
                    -\widehat{\dot\omega}_i\bigr]$,
so that 
\begin{align}
\widetilde{\sfT}_i(z_i,\widehat{\dot\omega}_i,\widehat{\omega}_i) = \proj_{[\widetilde a_i(z_i,\widehat{\dot\omega}_i,\widehat{\omega}_i),
         \widetilde b_i(z_i,\widehat{\dot\omega}_i,\widehat{\omega}_i)]}(0). 
\end{align}
Note that $\widetilde b_i - \widetilde a_i \geq 0$, so the projection is again well defined.  
Under ideal (i.e., error free) measurements,
we have that $\widetilde a_i(z_i, \widehat{\dot\omega},\widehat{\omega}) =a_i(z;x,d)$ and $\widetilde b_i(z_i,\widehat{\dot\omega},\widehat{\omega})
=b_i(z;x,d)$. Hence, the filter equation
\eqref{eq:local-dynamic-filter-z} coincides with the
continuous-time relaxation
\begin{equation}
 \epsilon\dot z=-z+\sfT(z;x,d).
 \label{eq:dynamic-proximal-filter}
\end{equation}
This establishes a proximal representation of the
proposed local filter. The following result connects
its equilibria, with $(x,d)$ held fixed, to optimal solutions of~\eqref{eq:static-qp}.

\begin{proposition}[Fixed-point characterization]
\label{prop:proximal-fixed-points}
For every fixed $(x,d)$, the solution set of
\eqref{eq:dual-problem} is the nonempty set
$\Fix(\sfT(\cdot;x))$.
Let $\cP(x,d)$ denote the set of optimizers
of~\eqref{eq:static-qp}. Then
\begin{equation}
 \cP(x,d)
 =\Fix(\sfT(\cdot;x))+\ker(\Gamma).
 \label{eq:qp-fixed-point-relation}
\end{equation}
Thus, every fixed point is a QP optimizer, and every
optimizer has a fixed-point representative producing
the same effective correction $\Gamma z$.
If $\Gamma\succ0$, then
$\Fix(\sfT(\cdot;x))=\{z^\star(x,d)\}$.
\end{proposition}

The proof is provided in the Appendix. For fixed $(x,d)$, the equilibrium set of
\eqref{eq:dynamic-proximal-filter} is
$\Fix(\sfT(\cdot;x))$.
By Proposition~\ref{prop:proximal-fixed-points},
every such equilibrium is an optimizer of
\eqref{eq:static-qp}.

In the following section, we analyze the performance of the proposed dynamic CBF filter. We also comment on power constraints. Then, Section~\ref{sec:emt_framework} will test the proposed filter on realistic EMT simulations, where~\eqref{eq:local-dynamic-filter-plant} is replaced by high-fidelity models for generators, converters, and network.  

\section{Analysis of the Dynamic CBF Filter}
\label{sec:filter-analysis}

In this section, we seek to bound the deviation $x(t)-x^\star(t)$, where 
$x(t)$ is a trajectory of~\eqref{eq:local-dynamic-filter} and  $x^\star(t)$ denotes
a trajectory of~\eqref{eq:ideal-closed-loop} when the ideal filter is used.
We consider two cases: 

\noindent \emph{i)} $\Gamma\succ0$, for which we bound
$\|z(t)-z^\star(x(t),d(t))\|_P$; and,

\noindent \emph{ii)} singular
$\Gamma\succeq0$, for which we instead bound
$\dist_P(z(t),\mathcal S(x(t),d(t)))$, where 
\begin{equation*}
 \mathcal S(x,d)
 :=\Fix\bigl(\sfT(\cdot;x,d)\bigr) =\{v\in\real^m:v=\sfT(v;x,d)\}.
\end{equation*}

We start by defining 
\begin{align*}
 L & :=\lambda_{\max}(K^{1/2}\Gamma K^{1/2}), \\
 q_K & :=\|I_m-K^{1/2}\Gamma K^{1/2}\|_2,
 \label{eq:analysis-gain-constants}
\end{align*}
which will be used throughout this section, and by 
modeling the frequency 
estimation errors. Define $e_\omega(t):=\hat\omega_{\cF}(t)-C x(t)$ and $e_{\dot\omega}(t):=\widehat{\dot\omega}_{\cF}(t)
                         -C\dot x(t)$. Assume that the errors are measurable and essentially bounded.  Using these errors, note that 
$\widetilde a_i(z_i,\widehat{\dot\omega}_i,\widehat{\omega}_i) = a_i(z;x,d)-k_i\eta_i$ and $\widetilde b_i(z_i,\widehat{\dot\omega}_i,\widehat{\omega}_i)  = b_i(z;x,d)-k_i\eta_i$, where $$
\eta:=a e_\omega+e_{\dot\omega}.$$ 
Let 
$e_{m,i} = \proj_{[\widetilde a_i(z_i,\widehat{\dot\omega}_i,\widehat{\omega}_i),
         \widetilde b_i(z_i,\widehat{\dot\omega}_i,\widehat{\omega}_i)]}(0) - \sfT_i(z;x,d)$ 
be the difference between the
implemented projection and $\sfT_i(z;x,d)$, so that we can write the dynamic CBF filter~\eqref{eq:local-dynamic-filter-z} as a perturbed version of~\eqref{eq:dynamic-proximal-filter} as  $\epsilon\dot z=-z+\sfT(z;x,d)+e_m$. Since projection of
zero onto a translated interval is Lipschitz with constant one, $|e_{m,i}| \leq k_i|\eta_i|$, $i\in\cF$, and 
\begin{equation}
\|e_m(t)\|_P\leq\|\eta(t)\|_K\leq\overline e_m,
\end{equation}
where $P:=K^{-1}$, $\|v\|_P=(v^\top Pv)^{1/2}$,
$\|v\|_K=(v^\top Kv)^{1/2}$, and $\overline e_m$ is a uniform
error bound.  In particular, $\|\eta\|_K\leq a\|e_\omega\|_K+\|e_{\dot\omega}\|_K$,
so bounds on the two estimation errors provide a bound on
$\overline e_m$.

By Proposition~\ref{prop:proximal-fixed-points}, th set $\mathcal S(x,d)$ is
nonempty and contained in $\cP(x,d)$. 
For each fixed $(x,d)$, non-expansiveness of the weighted proximal
operator~\cite[Sec.~2.3]{parikh2014proximal} gives
\begin{equation}
 \begin{aligned}
 \|\sfT(z;x,d)-\sfT(v;x,d)\|_P
 &\leq\|(I_m-K\Gamma)(z-v)\|_P\\
 &\leq q_K\|z-v\|_P
 \end{aligned}
 \label{eq:analysis-map-lipschitz}
\end{equation}
for all $z,v\in\real^m$, uniformly in $(x,d)$.
If $L<2$, then $q_K\leq1$, with $q_K<1$ when
$\Gamma\succ0$ and $q_K=1$ when $\Gamma$ is singular.
These properties are used in the tracking results below.

\subsection{Distance Between Ideal and Dynamic Filter Trajectories}

As  an intermediate step for our analysis, we seek to bound the deviation $z(t)-z^\star(x(t), d(t))$, where 
$z^\star(x(t), d(t))$ is the ideal filter. We start with the assumption that $\Gamma\succ0$.

\vspace{.1cm}

\noindent $\triangle$ \textbf{\emph{Case} $\Gamma\succ0$}. Define $\nu_\beta(t):=\|\frac{\textrm d}{\textrm{d} t }\underline\beta(x(t),d(t))\|_K$, with $\nu_\beta$ defined almost everywhere whenever $b$ is
locally absolutely continuous. Recall that the constraint box $\mathcal W(x,d)=[\underline\beta(x,d),
                    \underline\beta(x,d)+\Delta\one_m]$, $\Delta:=a(\overline\omega-\underline\omega)>0$,  
has fixed width. We have the following result. 

\vspace{.1cm}

\begin{proposition}[Filter error with $\Gamma\succ0$]
\label{prop:filter-pd-tracking}
Assume that $\Gamma\succ0$, and
fix $\epsilon>0$. Suppose that
\begin{equation}
 \lambda_{\max}(K^{1/2}\Gamma K^{1/2})<2.
 \label{eq:analysis-gain-condition}
\end{equation}
Let $(x,z)$ be an absolutely continuous solution
of~\eqref{eq:local-dynamic-filter} on $[0,T]$, with $d$
absolutely continuous and $e_\omega,e_{\dot\omega}$
measurable and essentially bounded on this interval.
Let $\overline e_m$ satisfy
$\|\eta(t)\|_K\leq\overline e_m$ for almost every $t\in[0,T]$. Then, the map $t\mapsto z^\star(x(t),d(t))$ is
absolutely continuous on $[0,T]$, and the error
$E_z(t):=\|z(t)-z^\star(x(t),d(t))\|_P$ satisfies
\begin{equation}
 \begin{aligned}
 E_z(t)\leq{}&\e^{- \frac{(1-q_K) t}{\epsilon}}E_z(0)\\
 &\hspace{-.7cm} +\int_0^t \e^{-\frac{(1-q_K)(t-s)}{\epsilon}}
     \left(\frac{\nu_\beta(s)}{1-q_K}
             +\frac{\|e_m(s)\|_P}{\epsilon}\right) \textrm{d} s , 
 \end{aligned}
 \label{eq:analysis-pd-tracking}
\end{equation}
for every $t\in[0,T]$.  \hfill $\Box$
\end{proposition}

The proof is in the Appendix. The proposition assumes that a solution $(x,z)$ exists on
$[0,T]$ and establishes a tracking bound along that solution.
When the estimation errors are measurable
and essentially bounded, existence and
uniqueness follow because the closed-loop vector field is
measurable in $t$ and globally Lipschitz in $(x,z)$.

\vspace{.1cm}

\noindent $\triangle$ \textbf{\emph{Case}}  $\Gamma\succeq0$. Under~\eqref{eq:analysis-gain-condition}, singularity of
$\Gamma$ gives $q_K=1$; therefore, Proposition~\ref{prop:filter-pd-tracking} does not apply. Instead, we study the distance to
$\mathcal S(x(t),d(t))$. Recall that
$\mathcal R:=\range(\Gamma)=\range(\Phi)$,
$\one_m\in\mathcal R$, and
$\underline\beta(x,d)\in\mathcal R$. We start with the following lemma. 

\begin{lemma}[Regularity and boundedness]
\label{lem:filter-set-regularity}
Let $\Gamma\succeq0$ and diagonal $K\succ0$ satisfy
\eqref{eq:analysis-gain-condition}. Suppose the constraint
boxes have the form $[b,b+\Delta\one_m]$, where
$b=\underline\beta(x,d)\in\mathcal R$, $\mathcal R:=\range(\Gamma)$, and 
$\Delta>0$ is fixed.
Then $\mathcal S(x,d)$ is a nonempty, compact, and convex
polyhedron, and there exists $0\leq\ell_S<\infty$ such that
\begin{equation}
 d_H^P\bigl(\mathcal S(x,d),
           \mathcal S(\widetilde x,\widetilde d)\bigr)
 \leq\ell_S
       \|\underline\beta(x,d)
          -\underline\beta(\widetilde x,\widetilde d)\|_K.
 \label{eq:analysis-set-lipschitz}
\end{equation}
For every fixed compact set of shifts $b\in\mathcal R$
and every bounded set of filter states, there exists a
uniform $\kappa\in(0,\infty)$ such that
\begin{equation}
 \dist_P(z,\mathcal S(x,d))
 \leq\kappa\|z-\sfT(z;x,d)\|_P.
 \label{eq:analysis-residual-error-bound}
\end{equation}

Moreover, consider an absolutely continuous solution on
$[0,T]$ of $\epsilon\dot z=-z+\sfT(z;x,d)+e_m$, with $|e_{m,i}|\leq k_i|\eta_i|$,  $\epsilon,T>0$, and measurable $\eta$. Suppose  $b \in \mathcal R$, $z(0)$ is in a bounded set, and $\|\eta(t)\|_\infty\leq\overline\eta<\Delta/2$ for almost every $t\in[0,T]$.
Then, the trajectory $z(t)$ remains in a bounded set. \hfill $\Box$
\end{lemma}

The proof of the lemma uses the polyhedral support function and
Hoffman's error bound~\cite{hoffman1952} and is given in the
Appendix. We then have the following result. 

\begin{proposition}[Filter error with $\Gamma\succeq0$]
\label{prop:filter-psd-tracking}
Assume that $\Gamma\succeq0$ is singular and that~\eqref{eq:analysis-gain-condition} holds. Let $(x,z)$ be an absolutely continuous solution
of~\eqref{eq:local-dynamic-filter} on $[0,T]$, with
$d$ absolutely continuous and $e_\omega,e_{\dot\omega}$
measurable and essentially bounded on this interval.
Assume that $\underline\beta(x(t),d(t))$ is in a compact subset of $\range(\Gamma)$ for every
$t\in[0,T]$. Suppose also that
$\|\eta(t)\|_\infty\leq\overline\eta<\frac{\Delta}{2}$ for almost every  $t\in[0,T]$.
Then,  $E_S(t):=\dist_P
   \bigl(z(t),\mathcal S(x(t),d(t))\bigr)$ is absolutely continuous on $[0,T]$ and, for
every $t\in[0,T]$,
\begin{equation}
 \begin{aligned}
 E_S(t)\leq & ~ \e^{-\frac{t}{2\kappa^2 \epsilon}}E_S(0)\\
 &+\int_0^t \e^{-\frac{t-s}{2\kappa^2 \epsilon}}
   \left(\ell_S\nu_\beta(s)
         +\frac{\|e_m(s)\|_P}{\epsilon}\right)ds,
 \end{aligned}
 \label{eq:analysis-psd-tracking}
\end{equation}
where $\ell_S$ is the constant in
\eqref{eq:analysis-set-lipschitz} and $\nu_\beta(t)
 =\left\|\frac{\textrm{d}}{\textrm{d} t}
   \underline\beta(x(t),d(t))\right\|_K$
is defined almost everywhere. \hfill $\Box$
\end{proposition}

The proof is postponed to the Appendix. We note that when $\Gamma$ is singular all QP optimizers $z^\star(x,d)$ have the same image under $\Gamma$ and, 
since $\ker(\Gamma)=\ker(G)$, the same $G z^\star(x,d)$.
For any $z^\star(x,d)\in\cP(x,d)$, choose a
nearest point to $z$ in $\mathcal S(x,d)$. Its difference
from $z^\star(x,d)$ is in $\ker(\Gamma)$, so
\begin{equation}
 \begin{aligned}
 \|\Gamma(z-z^\star)\|_2
 &\leq\|\Gamma K^{1/2}\|_2E_S,\\
 \|G(z-z^\star)\|_2
 &\leq\|G K^{1/2}\|_2E_S.
 \end{aligned}
 \label{eq:analysis-effective-action-tracking}
\end{equation}
If $\Gamma\succ0$, the two inequalities hold with $E_z$ in place of $E_S$.

\subsection{Distance Between State  Trajectories} 
\label{sec:input_constraints}

Recall that $x^\star(t)$ denotes
a trajectory of~\eqref{eq:ideal-closed-loop} when the ideal filter is used; i.e., of $\dot x^\star(t) = Ax^\star(t)+Bd(t)+G z^\star(x^\star(t),d(t))$ where $z^\star(x,d)$ is any optimizer of the ideal filter. To treat both cases for $\Gamma$, write
\begin{equation*}
 (E_{\rm f},\alpha,\ell):=
 \begin{cases}
 (E_z,\,1-q_K,\,(1-q_K)^{-1}), & \Gamma\succ0,\\
 (E_S,\,1/(2\kappa^2),\,\ell_S), & \Gamma\succeq0\text{ singular},
 \end{cases}
\end{equation*}
as in Propositions~\ref{prop:filter-pd-tracking} and~\ref{prop:filter-psd-tracking}, respectively, and define
\begin{subequations}
 \label{eq:analysis-plant-tracking-constants}
 \begin{align}
 g&:=\|G K^{1/2}\|_2,\\
 L_{\beta,x}&:=\|K^{1/2}C(aI+A)\|_2,\\
 L_{\beta,d}&:=\|K^{1/2}C B\|_2.
 \end{align}
\end{subequations}
Similarly to~\cite{dall2026local},  
we will use the notion of activation interval of the ideal filter, defined here as
\[
 \begin{aligned}
 \mathcal T_{\rm act}:=
 \bigl\{t\in[0,T]:\;&
G z^\star(x(t),d(t))\neq0\\
 &\text{or}~
G z^\star(x^\star(t),d(t))\neq0
 \bigr\},
 \end{aligned}
\]
Then, the following holds for $E_x(t) := \|x(t)-x^\star(t)\|_2$.  

\begin{proposition}[Distance from ideal state trajectory]
\label{prop:analysis-closed-loop-tracking}
Suppose that the assumptions of
Proposition~\ref{prop:filter-pd-tracking} or
Proposition~\ref{prop:filter-psd-tracking} hold on $[0,T]$.
Let $x^\star(t)$ and $x(t), z(t)$ be
solutions of~\eqref{eq:ideal-closed-loop} and~\eqref{eq:local-dynamic-filter}, respectively, on $[0,T]$ with the same $d(t)$. Assume that $x(t),x^\star(t)\in\mathcal X$ and
$d(t)\in\mathcal D$, where $\mathcal X,\mathcal D$
are compact sets. Suppose that $d(t)$ is
absolutely continuous  and $\overline\nu_d:=
 \operatorname*{ess\,sup}_{t\in[0,T]}
       \|\dot d(t)\|_2<\infty$, and $\|e_m(t)\|_P\leq\overline e_m$. Let $M_A\geq1$ and $\gamma_A>0$ satisfy $\|e^{At}\|_2\leq M_Ae^{-\gamma_A t}$,
  $t \geq 0$. If
\begin{equation}
 \frac{\alpha}{\epsilon}-\ell L_{\beta,x}g>0,
 \label{eq:analysis-closed-loop-timescale}
\end{equation}
then, for every $t\in[0,T]$,
\begin{equation}
 \begin{aligned}
 E_x(t)
 \leq{}&
 M_A \e^{-\gamma_A t+\zeta T_{\rm act}(t)}
       E_x(0)\\
 &\hspace{-.4cm} +\frac{M_Ag\, \e^{\zeta T_{\rm act}(t)}}
              {\alpha-\epsilon\ell L_{\beta,x}g}
 \Bigg[
   \epsilon E_{\rm f}(0)
   +\frac{1}{\gamma_A}
       \bigl(\epsilon\ell v_0+\overline e_m\bigr)
 \Bigg] , 
 \end{aligned}
 \label{eq:analysis-plant-state-bound}
\end{equation}
where $T_{\rm act}(t):=
 \int_0^t\mathbf{1}_{\mathcal T_{\rm act}}(s)\,ds$, $ \zeta:=M_Ag\ell L_{\beta,x}$, $M_\star<\infty$ is such that $\sup_{(x,d)\in\mathcal X\times\mathcal D}
       \|Ax+Bd+G z^\star(x,d)\|_2\leq M_\star$, and $v_0:=L_{\beta,x}M_\star
          +L_{\beta,d}\overline\nu_d$. \hfill $\Box$
\end{proposition}

A few remarks on
Proposition~\ref{prop:analysis-closed-loop-tracking} and, in particular, the bound~\eqref{eq:analysis-plant-state-bound}, 
are provided next.

\emph{(i) Approximation as $\epsilon\to0^+$.}
Fix $T>0$ and $\epsilon_0>0$ such that
$\epsilon_0\ell L_{\beta,x}g<\alpha$. Suppose the assumptions
of Proposition~\ref{prop:analysis-closed-loop-tracking}
hold on $[0,T]$ with common operating regions and constants
for all $0<\epsilon\leq\epsilon_0$. Assume that $E_x(0)=0$
and $E_{\rm f}(0)\leq\overline E_0$, where $\overline E_0$
is independent of $\epsilon$. Since $T_{\rm act}(t)\leq T$,
\begin{equation}
 \sup_{t\in[0,T]} E_x(t)
 \leq
 \frac{M_Ag\,\e^{\zeta T}}
      {\alpha-\epsilon_0\ell L_{\beta,x}g}
 \left[
   \epsilon\left(\overline E_0+\frac{\ell v_0}{\gamma_A}\right)
   +\frac{\overline e_m}{\gamma_A}
 \right].
\end{equation}
For $\overline e_m=0$, the trajectory deviation is
$O(\epsilon)$ uniformly on $[0,T]$. For a fixed
$\overline e_m>0$, the upper bound contains a term
proportional to $\overline e_m$ that persists as
$\epsilon\to0^+$.

\emph{(ii) Finite activation time.}
Both terms in
\eqref{eq:analysis-plant-state-bound}
are non-decreasing functions of $T_{\rm act}(t)$.
For a fixed $\epsilon$, suppose that
\[
 \overline T_{\rm act}
 :=\sup_{t\geq0}T_{\rm act}(t)<\infty.
\]
Then the proposition gives a finite upper bound
on $\sup_{t\geq0}E_x(t)$: replace
$T_{\rm act}(t)$ by $\overline T_{\rm act}$ and
$\e^{-\gamma_A t}$ by $1$ in
\eqref{eq:analysis-plant-state-bound}.
Finite filter activation time after a large disturbance, as illustrated in Fig.~\ref{fig:F_concept}, is consistent with the power systems 
operation in which primary control drives a
post-disturbance equilibrium within the
prescribed frequency limits, followed by secondary control driving the frequency to the nominal one~\cite{kundur1994power}. If $d(t)$ remains
constant after a disturbance and both closed-loop trajectories converge
to a post-disturbance operating point  with frequencies strictly
inside the prescribed limits, the CBF filter becomes inactive along both trajectories
after a finite time, implying
$\sup_{t\geq0}T_{\rm act}(t)<\infty$.

\emph{(iii) Frequency constraint violation.}
Suppose that $x^\star(t)\in\mathcal C$ for all
$t\in[0,T]$. Let $c_{\omega,i}^{\top}$ denote the
row of $C$ associated with bus $i\in\cF$.
For the reduced model in~\eqref{eq:local-dynamic-filter-plant}, one has that
\[
 |\omega_i(t)-\omega_i^\star(t)|
 \leq \|c_{\omega,i}\|_2 E_x(t).
\]
Since $\omega_i^\star(t)\in
[\underline\omega,\overline\omega]$, it follows that
\begin{align}
 \max\!\left\{
 \underline\omega-\omega_i(t),\,
 0,\,
 \omega_i(t)-\overline\omega
 \right\}
 \leq \|c_{\omega,i}\|_2 E_x(t) . 
\end{align}
This gives an explicit upper bound
on violation of either frequency limit at bus $i$.
Taking the supremum over $t\in[0,T]$ gives the
corresponding bound on the largest violation
over the specified horizon. Under the uniformity argument in~\emph{(ii)}, with
$x(0)=x^\star(0)$ and $\overline e_m=0$, the
frequency violation bound is $O(\epsilon)$ uniformly
on every fixed interval $[0,T]$ and therefore
converges to zero as $\epsilon\to0^+$. We will assess the frequency violation as function of $\epsilon$ in the numerical studies. 

\emph{(iv) Choice of time interval.}
Although stated on $[0,T]$, all results apply on any
interval $[t_0,t_0+T]$ on which their assumptions hold.
The corresponding estimates follow by replacing  $t$ with $t-t_0$, evaluating initial conditions at
$t_0$, and taking all time integrals, including the
accumulated activation time, from $t_0$.
The constants are those associated with the chosen
interval; uniformity with respect to $t_0$ requires
the assumptions and bounds to hold uniformly.

\subsection{Power Constraints} 
\label{sec:input_constraints}

Let
$\mathcal Z:=\prod_{i\in\cF}
[\underline z_i,\overline z_i]$
be a non-empty set of available powers for the filter; for example, for renewable-based systems, $[\underline z_i,\overline z_i]$ reflects the headroom allocated for the operation of the filter. The ideal QP~\eqref{eq:static-qp} can incorporate the
 constraint $p_{\sfs}\in\mathcal Z$; however, the resulting optimization problem  coupled across buses through $\Gamma$. Consequently,
component-wise clipping of the filter state does not,
in general, produce an optimizer of the power-constrained
QP. We therefore consider a local implementation that
clips the applied correction as: 
\begin{subequations}
\label{eq:local-dynamic-filter-constraints}
\begin{align}
 \dot x
 &=Ax+Bd+Gz^{\rm cl},
 \label{eq:local-dynamic-filter-plant-constraints}\\
 \epsilon\dot z_i
 &=-z_i+
 \widetilde{\sfT}_i
 (z_i,\widehat{\dot\omega}_i,\widehat{\omega}_i),
 \quad i\in\cF,
 \label{eq:local-dynamic-filter-z-constraints}\\
 z_i^{\rm cl}
 &=\proj_{[\underline z_i,\overline z_i]}(z_i),
 \quad i\in\cF,
 \label{eq:local-dynamic-filter-z-clipping}
\end{align}
\end{subequations}
where~\eqref{eq:local-dynamic-filter-z-clipping} performs a local saturation  of the filter output. This local implementation does not solve a QP with power constraints. The following analysis compares this implementation with
the original ideal closed loop~\eqref{eq:ideal-closed-loop} without saturation (i.e., with sufficient headroom for power correction).

To quantify the effect of the saturation, define the ``clipping error''
$s:=z^{\rm cl}-z$. The frequency-derivative measurement satisfies $\widehat{\dot\omega}_{\cF}
 =\varphi(x,d)+\Gamma z^{\rm cl}+e_{\dot\omega}$. Thus, with ideal measurements, the implemented map is $\sfT^{\rm cl}(z;x,d)
 :=
 \prox_{r(\cdot;x,d)}^{P}
       (z-K\Gamma z^{\rm cl})$.
Define
\[
 \begin{aligned}
 e_{\rm sat}
 &:=
 \sfT^{\rm cl}(z;x,d)-\sfT(z;x,d),\\
 e_m^{\rm cl}
 &:=
 \widetilde{\sfT}
 (z,\widehat{\dot\omega}_{\cF},\widehat{\omega}_{\cF})
 -\sfT^{\rm cl}(z;x,d), 
 \end{aligned}
\]
where $e_m^{\rm cl}$ accounts only for measurement errors. With $\eta:=ae_\omega+e_{\dot\omega}$, non-expansiveness of the proximal map
 gives
\begin{equation}
 \begin{aligned}
 \|e_m^{\rm cl}\|_P&\leq\|\eta\|_K,\\
 \|e_{\rm sat}\|_P
 &\leq\|K\Gamma s\|_P
 \leq L\|s\|_P,
 \end{aligned}
 \label{eq:clipping-map-errors}
\end{equation}
where
$L:=\lambda_{\max}(K^{1/2}\Gamma K^{1/2})$.
Therefore,~\eqref{eq:local-dynamic-filter-constraints}
can be re-written exactly as the following perturbed version of the ideal dynamic filter:
\begin{equation}
 \begin{aligned}
 \dot x
 &=Ax+Bd+G z+G s,\\
 \epsilon\dot z
 &=-z+\sfT(z;x,d)+e_{\rm sat}+e_m^{\rm cl}.
 \end{aligned}
 \label{eq:clipping-perturbed-system}
\end{equation}
Then, we have the following result.

\begin{proposition}[Trajectory bound under saturation]
\label{prop:clipped-actuation-tracking}
Let $(x,z)$ and $x^\star$ be absolutely continuous
solutions of~\eqref{eq:local-dynamic-filter-constraints}
and~\eqref{eq:ideal-closed-loop}, respectively, on $[0,T]$,
with the same $d$.
Suppose the assumptions
of Proposition~\ref{prop:analysis-closed-loop-tracking}
hold for $x(t)$. Assume
$L<2$ and $\epsilon\ell L_{\beta,x}g<\alpha$. Suppose that, almost everywhere on $[0,T]$,
\begin{equation}
 \|s(t)\|_P\leq\overline s,
 \qquad
 \|e_m^{\rm cl}(t)\|_P\leq\overline e_m.
 \label{eq:clipping-error-bounds}
\end{equation}

For $\Gamma\succ0$, take $\alpha=1-q_K$ and the
corresponding sensitivity constant $\ell$.
Then, for every $t\in[0,T]$,
 \begin{align}
  \label{eq:clipping-plant-bound}
 E_x(t)\leq &
 M_Ae^{-\gamma_A t+\zeta T_{\rm act}(t)}E_x(0)\\
 & \hspace{-.8cm} +
 \frac{M_Ag\,e^{\zeta T_{\rm act}(t)}}
      {\alpha-\epsilon\ell L_{\beta,x}g}
 \left[
   \epsilon E_z(0)
   +\frac{\epsilon\ell v_0+\overline e_m
                    +(\alpha+L)\overline s}{\gamma_A}
 \right]. \nonumber 
 \end{align}

\begin{figure*}[!t]
    \centering
    \includegraphics[width=\textwidth]{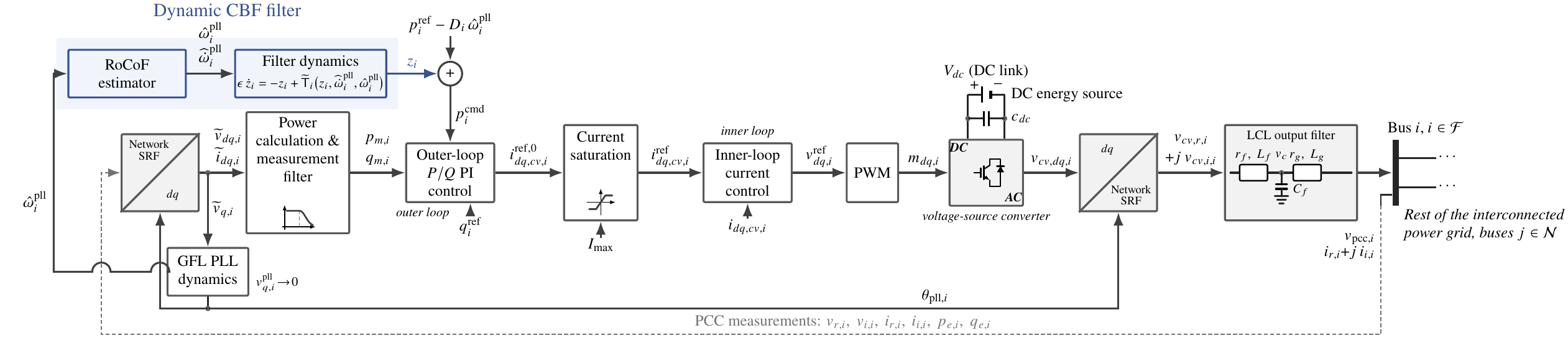}
    \caption{Higher-fidelity $dq0$ EMT representation of the GFL IBR with the
    proposed dynamic local safety filter. Electrical quantities and controllers are explained in  Appendix \ref{sec:inverterEMT}.}
    \label{fig:emt_plant}
\end{figure*}

For singular $\Gamma\succeq0$, suppose additionally that
the residual error bound of
Lemma~\ref{lem:filter-set-regularity} holds uniformly.
Then~\eqref{eq:clipping-plant-bound} holds with
$E_z(0)$ replaced by $E_S(0)$,
$\alpha=\sigma$, and $\ell=\ell_S$. \hfill $\Box$
\end{proposition}

The proof follows steps similar to the ones of Proposition~\ref{prop:analysis-closed-loop-tracking}, and it is omitted to avoid repetitive arguments. Proposition~\ref{prop:clipped-actuation-tracking}
assumes bounded difference between $z$ and 
applied correction $z^{\rm cl}$ on the interval considered;
it does not exclude transient windup. Persistent windup is not expected for contingencies
for which primary control brings back frequencies
strictly inside the prescribed limits. This will be shown in the EMT validation presented in the ensuing section.

\section{EMT Validation}
\label{sec:emt_framework}

While the filter design is based on a reduced power network model, we evaluate the performance of the proposed filters
using the toolbox
\textsc{PowerSimulationsDynamics.jl} (PSID.jl)
~\cite{lara2023powersimulationsdynamics}%
\footnote{\url{https://github.com/NREL-Sienna/PowerSimulationsDynamics.jl/releases}},
with models, data and parameters available online%
\footnote{\url{https://github.com/DamolaMicheal/SafetyFiltersTestbedEMT}}.
The EMT model includes GENROU-type generator models with shaft,
excitation, governor, and power-system-stabilizer controls;
dynamic-$\pi$ transmission lines; ZIP loads retaining
only the constant-power component; and GFL converter
dynamics comprising an SRF-PLL, outer P/Q PI control,
inner current control, and an LCL filter.
The GFL IBRs estimate
local frequency through a phase-locked loop (PLL) and
track active-power setpoints through finite-bandwidth
controls.

For higher-fidelity validation, the proposed dynamic local
safety filter~\eqref{eq:local-dynamic-filter} is implemented
in the balanced synchronous-frame $dq0$ EMT model for the GFL inverter shown in
Fig.~\ref{fig:emt_plant}. The implementation of the filter required a modification of the dynamic model in PSID.jl; these modifications are explained in detail in the Appendix \ref{sec:inverterEMT}.

\begin{figure*}[!t]
    \centering
    \includegraphics[
        width=\textwidth,
        height=0.82\textheight,
        keepaspectratio
    ]{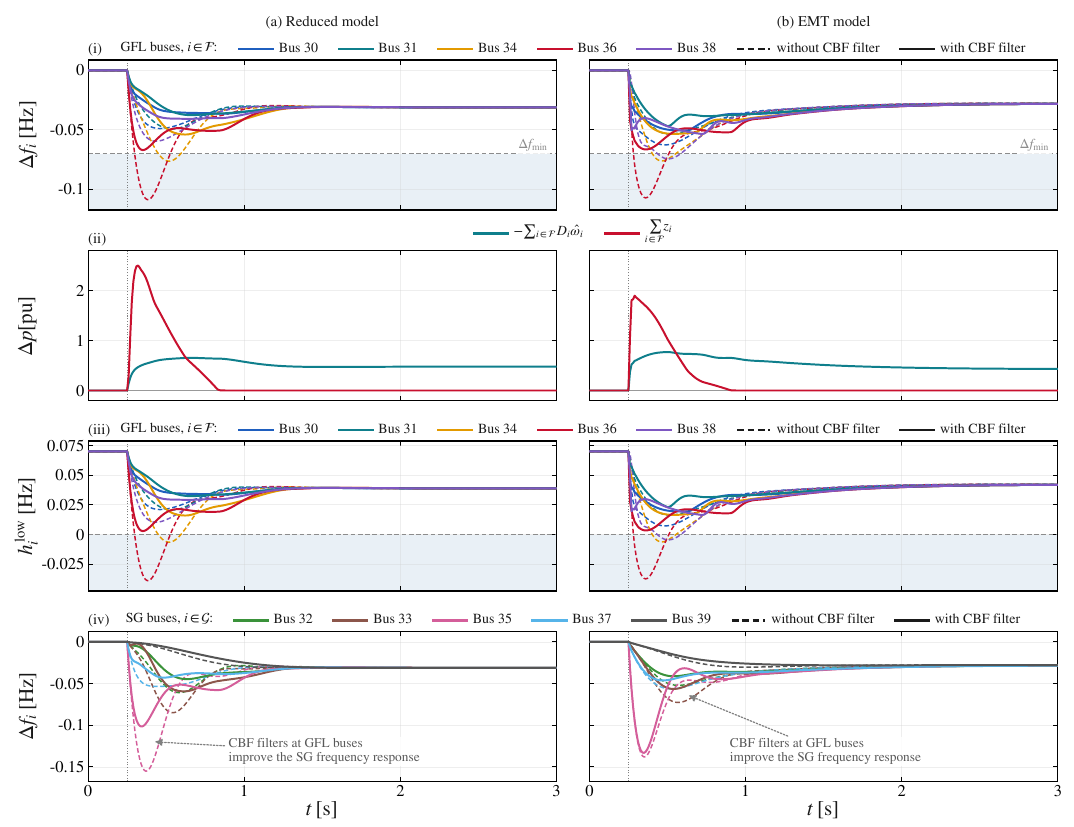}
    \caption{Underfrequency responses of the reduced-order
    model (left) and detailed $dq0$-EMT model (right)
    following a sudden load increase, for
    $|\mathcal F|=|\mathcal G|=5$.
    Panels show (i) GFL frequency deviations,
    (ii) aggregate primary-control and CBF active-power
    deviations,  (iii) GFL underfrequency safety margins, and
    (iv) SG frequency deviations.
    Solid and dashed trajectories correspond to
    operation with and without dynamic local CBF safety
    filters, respectively.}
    \label{fig:underfrequency_event}

    \vspace{-.4cm}
    
\end{figure*}

\begin{figure}[!t]
    \centering
    \includegraphics[
        width=\columnwidth,
        keepaspectratio
    ]{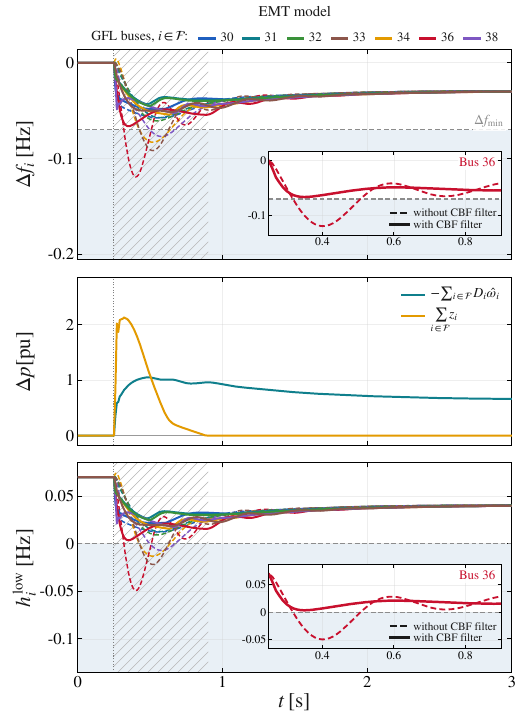}
    \caption{Response of the EMT model
    following a sudden load increase, with a dynamic
    local safety filter implemented at each GFL bus, for  $|\mathcal F|>|\mathcal G|$.}
    \label{fig:underfrequency_event_high_gfl}
    \vspace{-.4cm}
\end{figure}

We evaluate the proposed dynamic CBF  filter on a
modified IEEE 39-bus system with a nominal frequency of
$60$~Hz. The objective of the studies is threefold: \emph{(i)} compare reduced-order and detailed
$dq0$-EMT responses to sudden load increases, so that to verify the design pipeline. \emph{(ii)} Examine
frequency containment under different generator and GFL allocations
and GFL active-power saturation. \emph{(iii)} Assess the sensitivity to
the filter time constant $\epsilon$. The first objective is particularly important, as we use a reduce model as a design model, and EMT as a validation model. Two low-inertia
configurations are considered, with the same
operating conditions and disturbances used for the corresponding model comparisons. In the experiments, the dirty derivative time constant satisfies
$\tau_{d,i}<\epsilon$.

\vspace{.2cm}

\noindent \textbf{Scenario 1: Network configuration with
$|\cF|\leq|\cG|$.}
We first consider $\cF=\{30,31,34,36,38\}$ and
$\cG=\{32,33,35,37,39\}$, with bus~39 as the angle reference. In this case, the matrix $\Gamma$  is positive definite. 
At $t=0.25$~s, sudden load increases of $300$~MW, $150$~MW,
and $75$~MW are applied at buses~35, 2, and~29,
respectively. The prescribed GFL frequency-safety band is
$\Delta f_i\in[-0.07,0.07]$~Hz.
Figure~\ref{fig:underfrequency_event} compares the
reduced-order and $dq0$-EMT responses. Without the safety
filter, several GFL frequencies fall below the prescribed
lower limit, and the corresponding barrier functions become
negative. With the local filters enabled, the GFL frequencies
remain within the safety band, and the barrier functions
remain nonnegative in both models. Differences in the
frequency nadirs and short-time transients are consistent
with the additional PLL, converter-control, and
electromagnetic dynamics represented in the EMT model. The aggregate GFL power responses show the contributions
of the nominal primary control,
$-\sum_{i\in\cF}D_i\hat{\omega}_i$, and the CBF correction,
$\sum_{i\in\cF}z_i$, as defined in
\eqref{eq:supervisory-power-command}.
The corrective action increases during the underfrequency
transient and returns toward zero as the nominal GFL
response provides sufficient frequency support.
Although the safety filters act only at GFL buses, the
additional active-power support also propagate to reduce the SG
frequency excursions, showing the effect of the local
filters on the wider network.

\vspace{.2cm}

\noindent \textbf{Scenario 2: Network configuration with
$|\cF|>|\cG|$.}
We next replace the synchronous generators at buses~32 and~33 with GFL resources,
giving $\cF=\{30,31,32,33,34,36,38\}$ and
$\cG=\{35,37,39\}$, so that $|\cF|=7>|\cG|=3$. In this case, the matrix $\Gamma$  is positive semi-definite and singular.
The disturbance and frequency-safety limits remain unchanged
from Scenario~1. By Proposition~\ref{prop:gamma}, this
configuration makes $\Gamma$ singular. Although the ideal QP
admits nonunique optimal corrections, these produce the same
effective input to the plant, as discussed in
Section~\ref{sec:filter-analysis}. Figure~\ref{fig:underfrequency_event_high_gfl} shows the
corresponding $dq0$-EMT response. As in Scenario~1, the
GFL frequencies without CBF intervention violate the lower safety limit,
whereas the local filters maintain the prescribed bounds,
with nonnegative barrier functions. The aggregate CBF correction increases following the
disturbance and gradually returns toward zero as the
frequencies recover, with primary frequency support
remaining active. These results show that the dynamic
filter remains locally implementable irrespective of
the SG--GFL allocation.

\vspace{.2cm}

\begin{figure}[!t]
    \centering
    \includegraphics[
        width=\columnwidth,
        keepaspectratio
    ]{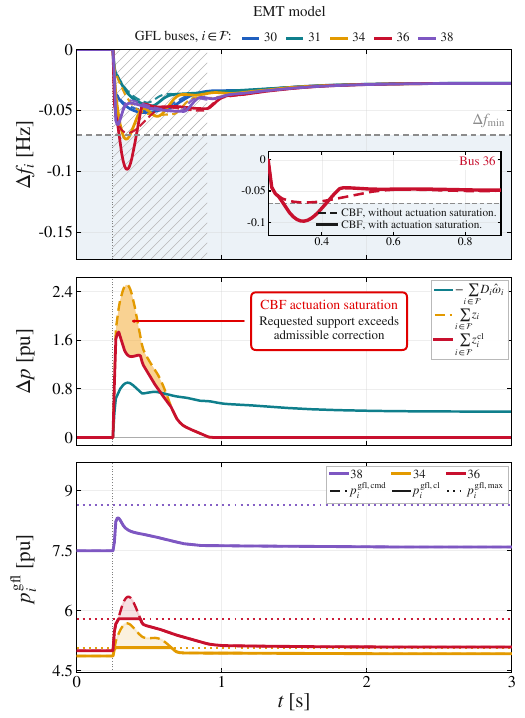}
    \caption{Effect of GFL active-power actuation limits on the EMT response
    to a sudden load increase with the dynamic CBF filter at each GFL bus.}
    \label{fig:emt_cbf_actuation_limits}

        \vspace{-.4cm}
        
\end{figure}

\noindent \textbf{Scenario 3: GFL active-power saturation.}
Using the system configuration and disturbance of Scenario~1,
we examine the effect of GFL active-power saturation,
implemented through the local projection in
\eqref{eq:local-dynamic-filter-z-clipping}.
Figure~\ref{fig:emt_cbf_actuation_limits} compares the
EMT responses with and without saturation. Following the load increase, the active-power corrections
requested by the CBF safety filters at buses~34 and~36
exceed the available inverter power limit, causing the applied
corrections to saturate. The aggregate
corrective power is reduced, and the frequency at bus~36
falls below the prescribed lower limit, unlike the
unconstrained actuation case. As the frequencies recover, the
corrective action returns toward zero. These results explain the effect of actuation saturation
analyzed in Proposition~\ref{prop:clipped-actuation-tracking}
and the need for sufficient GFL responsive reserves to
maintain the prescribed frequency-safety limits.

\vspace{.2cm}

\begin{figure}[!t]
    \centering
    \includegraphics[
        width=\columnwidth,
        keepaspectratio
    ]{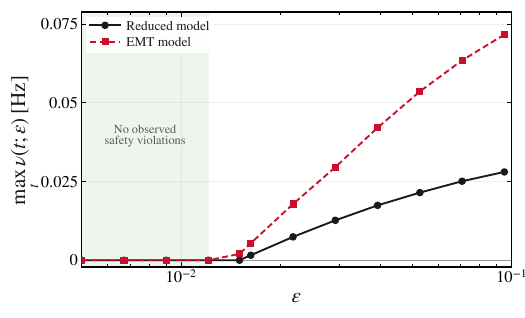}
    \caption{Maximum GFL underfrequency violation
    $\max_t\nu(t;\epsilon)$ versus the dynamic-filter
    time constant $\epsilon$ in the reduced-order and EMT models. The shaded region indicates
    no observed safety violations, $\nu(t;\epsilon):=\max\{0,\max_{i\in\cF}[\Delta f_{\min}-\Delta f_i(t;\epsilon)]\}$.}
    \label{fig:epsilon_sensitivity}

    \vspace{-.4cm}
    
\end{figure}

\noindent \textbf{Dynamic Safety-Filter Sensitivity.}
\label{subsec:sensitivity_analysis}
We examine the effect of the safety-filter time constant
$\epsilon$. As discussed in Section~\ref{sec:dynamic_local_filter},
$\epsilon$ determines how quickly the dynamic filter tracks
the ideal filter~\cite{dall2026local}. Smaller values allow faster corrective responses,
whereas larger values introduce greater lag during frequency
transients. Figure~\ref{fig:epsilon_sensitivity} shows the peak GFL
underfrequency violation $\max_t\nu(t;\epsilon)$ for the
reduced-order and EMT models. Neither model exhibits
safety violations for the tested values of $\epsilon$ up to
approximately $0.012$~s. As $\epsilon$ increases,
violations emerge and grow in magnitude, with the EMT
model exhibiting an earlier onset and greater sensitivity
than the reduced model.

\section{Conclusion and Future Work}
\label{sec:conclusion}
This paper presented dynamic local CBF safety filters for transient frequency containment in power transmission systems with grid-following inverters (GFL). Starting from the ideal network-coupled CBF quadratic program, a dual proximal formulation and two-time-scale design lead to local dynamic filters that use only frequency and RoCoF estimates. We established frequency-safety guarantees for the ideal filter and derived trajectory-error bounds for the local implementation, accounting for unique and nonunique optimal corrections, measurement errors, and inverter power saturation. Studies on the IEEE 39-bus system, including detailed EMT simulations, showed the effectiveness of the filters in containing frequency excursions. The results also show how filter time scale and active-power saturation can lead to frequency-limit violations. Future work will extend the safety filters to address frequency and voltage,  and investigate sampled-data implementation under measurement and power-actuation delays.

\appendix

\subsection{GFL inverter with safety filter in the $dq0$ EMT model}
\label{sec:inverterEMT}

At each controlled GFL bus, the SRF-PLL estimates the PCC voltage angle
and aligns the converter $dq$ control frame with the grid-voltage
phasor. Near synchronization, this alignment gives
$v_{q,i}^{\rm pll}\approx0$ and places the PCC voltage predominantly on the
$d$ axis, so active-power regulation is effected primarily through the
$d$-axis current. The PLL also provides the local frequency measurement
used to obtain the RoCoF estimate. These local signals provide the
feedback to the dynamic CBF filter, as shown in
Fig.~\ref{fig:emt_plant}. The following derivation describes
this measurement-to-actuation path within the EMT  model.

\noindent\emph{1) SRF-PLL and local frequency measurement:}
For each $i\in\cF$, let
$v_{{\rm pcc},i}:=[v_{r,i},\,v_{i,i}]^\top
=|v_i|[\cos\theta_{v,i},\,\sin\theta_{v,i}]^\top\in\real^2$
denote the PCC voltage in the common network real--imaginary reference
frame, where $v_{r,i}$ and $v_{i,i}$ are its real and imaginary
components, and $|v_i|$ and $\theta_{v,i}$ are its magnitude and
phase angle, respectively. Define the rotation matrix $R(\theta):=
\begin{bmatrix}
\cos\theta & -\sin\theta\\
\sin\theta & \cos\theta
\end{bmatrix}$.
The SRF-PLL~\cite{kaura1997operation} maps the PCC voltage
into its local $dq$ frame, filters the resulting components, and
estimates the local frequency through
\begin{subequations}
\label{eq:emt_pll}
\begin{align}
\widetilde v_{dq,i}
&=
R(-\theta_{{\rm pll},i})v_{{\rm pcc},i},
\label{eq:emt_pll_transform}
\\
\dot v_{dq,i}^{\rm pll}
&=
\omega_{{\rm lp},i}
\left(
\widetilde v_{dq,i}-v_{dq,i}^{\rm pll}
\right),
\label{eq:emt_pll_filter}
\\
e_{\theta,i}
&=
\operatorname{atan2}
\left(
v_{q,i}^{\rm pll},v_{d,i}^{\rm pll}
\right),
\qquad
\dot\eta_{{\rm pll},i}
=
e_{\theta,i},
\label{eq:emt_pll_error}
\\
\omega_{{\rm pll},i}^{\rm pu}
&=
1+
k_{p,{\rm pll},i}e_{\theta,i}
+
k_{i,{\rm pll},i}\eta_{{\rm pll},i},
\label{eq:emt_pll_frequency}
\\
\dot\theta_{{\rm pll},i}
&=
\Omega_b
\left(
\omega_{{\rm pll},i}^{\rm pu}
-\omega_{\rm sys}
\right),
\label{eq:emt_pll_angle}
\end{align}
\end{subequations}
where $\widetilde v_{dq,i}:=
[\widetilde v_{d,i},\,\widetilde v_{q,i}]^\top$ denotes the
pre-filter PCC voltage in the PLL-aligned $dq$ frame and
$v_{dq,i}^{\rm pll}:=
[v_{d,i}^{\rm pll},\,v_{q,i}^{\rm pll}]^\top$ its filtered value.
The states $\theta_{{\rm pll},i}$ and $\eta_{{\rm pll},i}$
denote the PLL angle relative to the network synchronous frame
and the PLL PI integrator state, respectively.
The parameters $\omega_{{\rm lp},i}$,
$k_{p,{\rm pll},i}$, and $k_{i,{\rm pll},i}$ denote the PLL
voltage-filter bandwidth and PI gains, respectively;
$\omega_{\rm sys}$ is the per-unit speed of the network synchronous
frame, and $\Omega_b=2\pi f_b$ is the base electrical radian frequency,
with $\Omega_b=2\pi60$~rad/s for the $60$~Hz system considered here.
For a constant grid frequency, synchronization gives
$e_{\theta,i}\rightarrow0$, giving
$v_{q,i}^{\rm pll}\rightarrow0$ and
$v_{d,i}^{\rm pll}\rightarrow\|v_{{\rm pcc},i}\|_2$. During frequency transients, the finite bandwidths of the
PLL and RoCoF estimator can introduce estimation errors,
whose effects on the dynamic safety filter are analyzed
in Section~\ref{sec:filter-analysis}. The PLL output $\omega_{{\rm pll},i}^{\rm pu}$ denotes the per-unit
estimate of the local PCC frequency at GFL bus $i$. The corresponding
frequency-deviation measurement used by the safety filter is $\hat\omega_i^{\rm pll}
:=
\Omega_b(\omega_{{\rm pll},i}^{\rm pu}-1)$.
As in~\eqref{eq:local_rocof_estimator},
the RoCoF estimate is obtained from $\hat\omega_i^{\rm pll}$ using the dirty derivative:
\begin{equation}
\tau_{d,i}\dot\chi_i
=
-\chi_i+\hat\omega_i^{\rm pll},
\qquad
\widehat{\dot\omega}_i^{\rm pll}
=
\frac{\hat\omega_i^{\rm pll}-\chi_i}{\tau_{d,i}}.
\label{eq:emt_rocof_estimator}
\end{equation}

\noindent\emph{2) GFL converter control and safety-filter interface:}
The GFL outer loop computes the instantaneous terminal active
and reactive powers from the network-frame voltage and current
as $p_{e,i}=v_{r,i}i_{r,i}+v_{i,i}i_{i,i}$ and
$q_{e,i}=v_{i,i}i_{r,i}-v_{r,i}i_{i,i}$.
The PLL frequency estimate $\hat\omega_i^{\rm pll}$
and RoCoF estimate $\widehat{\dot\omega}_i^{\rm pll}$
provide the local measurements for the dynamic safety filter~\eqref{eq:local-dynamic-filter-z}.
The resulting correction $z_i$  modifies the commanded active-power setpoint according to (see \eqref{eq:supervisory-power-command}):
\begin{equation}
p_i^{\rm gfl}
=
\bar p_i^{\rm gfl}
-D_i\hat\omega_i^{\rm pll}
+z_i.
\label{eq:emt_active_power_command}
\end{equation}
The power quantities
in \eqref{eq:emt_active_power_command} are expressed on
a common inverter base. When GFL active-power limits
are imposed, $z_i$ is replaced by the saturated correction
$z_i^{\rm cl}$ defined in
\eqref{eq:local-dynamic-filter-z-clipping}.
Let $p_{m,i},q_{m,i}\in\real$ denote the filtered power
measurements and $\sigma_{p,i},\sigma_{q,i}\in\real$
the outer-loop PI states. Their dynamics are
\begin{subequations}
\label{eq:emt_outer_control}
\begin{align}
\dot p_{m,i}
&=
\omega_{z,i}(p_{e,i}-p_{m,i}),
&
\dot\sigma_{p,i}
&=
p_i^{\rm gfl}-p_{m,i},
\label{eq:emt_outer_active}
\\
\dot q_{m,i}
&=
\omega_{f,i}(q_{e,i}-q_{m,i}),
&
\dot\sigma_{q,i}
&=
q_i^{\rm ref}-q_{m,i},
\label{eq:emt_outer_reactive}
\end{align}
\end{subequations}
where $\omega_{z,i}$ and $\omega_{f,i}$ are the angular cutoff
frequencies of the active- and reactive-power measurement filters,
respectively. The current references
in the PLL-aligned $dq$ frame are
\begin{subequations}
\label{eq:emt_current_references}
\begin{align}
i_{d,{\rm cv},i}^{\rm ref}
&=
k_{pp,i}(p_i^{\rm gfl}-p_{m,i})
+
k_{pi,i}\sigma_{p,i},
\label{eq:emt_id_reference}
\\
i_{q,{\rm cv},i}^{\rm ref}
&=
-k_{qp,i}(q_i^{\rm ref}-q_{m,i})
-k_{qi,i}\sigma_{q,i},
\label{eq:emt_iq_reference}
\end{align}
\end{subequations}
where $k_{pp,i}$ and $k_{pi,i}$ are the active-power
PI gains, while $k_{qp,i}$ and $k_{qi,i}$ are the
reactive-power PI gains. Notice from~\eqref{eq:emt_active_power_command}--\eqref{eq:emt_id_reference}
that $z_i$ acts through $p_i^{\rm gfl}$ to modify
$i_{d,{\rm cv},i}^{\rm ref}$. 
As shown in Fig.~\ref{fig:emt_plant}, the resulting $dq$
current references are subject to the converter current
limit $I_{\max}$ before being tracked by the inner current loop. The inner PI controller uses $dq$ decoupling and voltage
feedforward to generate the converter voltage reference,
which is applied through the PWM stage to the DC-link-fed
VSC~\cite{yazdani2010voltage}. The LCL filter captures
the dynamics of the converter-side current, filter-capacitor
voltage, and grid-side current injected at the PCC. Thus, within the EMT plant, the supervisory local safety filter acts through
the outer active-power loop, with the downstream converter dynamics
represented by the detailed GFL model. We refer readers to
Appendix~B of~\cite{ajeyemi2026reachability}
for the detailed dynamics and associated equations of the GFL
converter model.

\subsection{Proof of Proposition~\ref{prop:ideal-invariance}}

We first establish feasibility of~\eqref{eq:static-qp}
for every $(x,d)$. Since
$\range(\Gamma)=\range(\Phi)$ and
$\Phi\one_{n_{\cG}}=\one_m$, we have $C x,\ C(Ax+Bd),\ \one_m
 \in\range(\Gamma)$.
Consequently, $\underline\beta(x,d)\in\range(\Gamma)$.
The choice
$v=\Gamma^\dagger\underline\beta(x,d)$
therefore satisfies $\Gamma v=\underline\beta(x,d)
 \leq\overline\beta(x,d)$,
because
$\overline\beta-\underline\beta
=a(\overline\omega-\underline\omega)\one_m>0$.
Thus, the QP is feasible on the entire state space,
which contains $\mathcal C$.

Let $\mathcal R:=\range(\Gamma)=\range(\Phi)$ and
$b:=\underline\beta(x,d)$.
Since $C x$, $C(Ax + Bd)$, and $\one_m$ belong to
$\mathcal R$, so does $b$. Thus, $\Gamma^\dagger b$ is feasible
for~\eqref{eq:static-qp}. Define the  compact convex set
$\mathcal K:=
 \mathcal R\cap
 [\,\zero_m,\ a(\overline\omega-\underline\omega)\one_m\,]$. In the variable $w:=\Gamma v$, the QP has
the unique optimizer $w^\star(b)=b+\proj_{\mathcal K}^{\Gamma^\dagger}(-b)$,
where the projection uses the positive-definite metric
induced by $\Gamma^\dagger$ on $\mathcal R$.
Non-expansiveness of the projection  implies that
$w^\star$ is globally Lipschitz in $b$.
Furthermore, $\ker(\Gamma)=\ker (G)$, so every optimizer
induces the same plant input
$G\Gamma^\dagger w^\star(b)$.
The resulting vector field is globally Lipschitz in $x$,
with linear growth in $(x,d)$, yielding a unique global
absolutely continuous state trajectory.

QP feasibility gives, almost everywhere, $\dot{\underline{h}}_i\geq-a \underline{h}_i$ and 
$\dot{\overline{h}}_i\geq-a \overline{h}_i$. Hence, for all $t\geq0$,
\[
 \begin{aligned}
 \underline{h}_i(x(t))
 &\geq e^{-at}\underline{h}_i(x(0)),\\
 \overline{h}_i(x(t))
 &\geq e^{-at}\overline{h}_i(x(0)).
 \end{aligned}
\]
Both remain nonnegative when $x(0)\in\mathcal C$.

\subsection{Proof of Proposition~\ref{prop:proximal-fixed-points}}

Fix $(x,d)$ and suppress this dependence in $r$ and $g$
for notational simplicity.
Proximal optimality gives
\[
 z=\sfT(z;x,d)
 \quad\Longleftrightarrow\quad
 0\in\Gamma z+\partial r(z),
\]
which characterizes the minimizers of~\eqref{eq:dual-problem}.

The QP~\eqref{eq:static-qp} is strictly feasible and has
a finite optimal value. 
The primal optimum is attained:  equivalently, one
minimizes $\frac12w^\top\Gamma^\dagger w$ over the
nonempty compact set
$\mathcal W(x,d)\cap\range(\Gamma)$.
An optimal $w$ gives the primal optimizer
$\Gamma^\dagger w$. By Slater's theorem, the maximum
of $g$ is attained and equals the QP optimal
value~\cite[Sec.~5.2.3]{vandenberghe2004convex}.
Equivalently,~\eqref{eq:dual-problem} has a minimizer
$z^\star$. The optimality condition gives
$z^\star=\sfT(z^\star;x,d)$, proving that
$\Fix(\sfT(\cdot;x,d))$ is nonempty.

Let $v^\star$ be any optimizer of
\eqref{eq:static-qp} and let
$z^\star\in\Fix(\sfT(\cdot;x,d))$.
Strong duality yields
\[
 \begin{aligned}
 0
 &=\frac12(v^\star)^\top\Gamma v^\star
      -g(z^\star)\\
 &=\frac12(v^\star-z^\star)^\top
          \Gamma(v^\star-z^\star) +r(z^\star)+(z^\star)^\top\Gamma v^\star.
 \end{aligned}
\]
The quadratic term is nonnegative because $\Gamma\succeq0$.
The remaining term is nonnegative by the definition of $r$
and the feasibility of $v^\star$.
Consequently, $v^\star-z^\star\in\ker(\Gamma)$.
Thus, $z^\star$ is primal feasible and has the same
objective value as $v^\star$.

Since this holds for every primal--dual optimal pair,
and adding an element of $\ker(\Gamma)$ preserves
primal feasibility and cost,
\eqref{eq:qp-fixed-point-relation} follows.
When $\Gamma\succ0$, the objective of the QP is strictly
convex and $\ker(\Gamma)=\{0\}$, proving the final claim. Therefore, every dual optimizer is also a primal optimizer of this particular QP.

\subsection{Proof of Proposition~\ref{prop:filter-pd-tracking}}

The gain condition and $\Gamma\succ0$ imply $q_K<1$.
For brevity, write
$b(t):=\underline\beta(x(t),d(t))$ and
$z^\star(t):=z^\star(x(t),d(t))$.

Since the constraint box has fixed width, changing from $b(s)$ to $b(t)$ shifts both endpoints
of the interval by $k_i(b_i(t)-b_i(s))$.
The scalar projection~\eqref{eq:scalar-proximal-map}
therefore gives, for every $v\in\real^m$ and $s,t\in[0,T]$,
\[
 \|\sfT(v;x(t),d(t))-\sfT(v;x(s),d(s))\|_P
 \leq\|b(t)-b(s)\|_K.
\]
Combining this estimate
with~\eqref{eq:analysis-map-lipschitz} and the fixed-point identity yields $(1-q_K)\|z^\star(t)-z^\star(s)\|_P
 \leq\|b(t)-b(s)\|_K$.
Since $\underline\beta$ is affine and $x,d$ are absolutely
continuous on $[0,T]$, so is $b$. The previous inequality
implies that $p_\sfs^\star(\cdot)$ is absolutely continuous
on $[0,T]$, with
\[
 \|\dot z^\star(t)\|_P
 \leq\frac{\nu_\beta(t)}{1-q_K}
 \quad\text{for almost every }t\in[0,T].
\]

Recall that $E_z(t):=\|z(t)-z^\star(t)\|_P$. The function $E_z$ is absolutely continuous; at almost every time for which $E_z(t)>0$, using the perturbed dynamics
$\epsilon\dot z=-z+\sfT(z;x,d)+e_m$,
the identity
$\sfT(z^\star(t);x(t),d(t))=z^\star(t)$,
and~\eqref{eq:analysis-map-lipschitz}, gives
\[
 \dot E_z(t)
 \leq-\frac{1-q_K}{\epsilon}E_z(t)
      +\|\dot z^\star(t)\|_P
      +\frac{\|e_m(t)\|_P}{\epsilon}.
\]
Since $E_z$ is nonnegative and absolutely continuous,
$\dot E_z=0$ almost everywhere on $\{t:E_z(t)=0\}$.
Hence, the same inequality holds almost everywhere
on $[0,T]$.
Applying Grönwall's inequality yields~\eqref{eq:analysis-pd-tracking}.

\subsection{Proof of Lemma~\ref{lem:filter-set-regularity}}
\label{app:filter-set-regularity}

Write $b=\underline\beta(x,d)$, $Q=\Gamma^{1/2}$, and
$r_0(z)=\Delta\sum_{i\in\cF}\max\{-z_i,0\}$ for brevity.
The dual objective is $F_b(z)=\tfrac12\|Qz\|_2^2-b^\top z+r_0(z)$,
with minimizer set $\mathcal S_b:=\mathcal S(x,d)$.
Let $Q^\dagger$ denote the Moore-Penrose inverse of $Q$. The three claims of the lemma are proved next. 

\emph{Compactness and variation of the minimizer set.}
For $\xi\in\mathcal R$, define $\psi(\xi):=\min_{Qz=\xi}r_0(z)$ and $Z(\xi):=\argmin_{Qz=\xi}r_0(z)$.   For every $\xi\in\mathcal R$, the problem
$\min_{Qz=\xi}r_0(z)$ is feasible and has a finite optimal
value. By~\cite[Cor.~19.3.1, p.~175]{rockafellar1970},
the minimum is attained and $\psi$ is a polyhedral convex
function. On $\mathcal R$, it can therefore be represented as the
maximum of finitely many affine functions
\cite[p.~172]{rockafellar1970}, which implies global
Lipschitz continuity on $\mathcal R$. Denote a Lipschitz constant by $L_\psi$. Since $\one_m\in\mathcal R$,
\begin{equation}
 \|z\|_1
 \leq a_1\|Qz\|_2+\frac{2r_0(z)}{\Delta},
 \label{eq:appendix-fiber-bound}
\end{equation}
with $a_1:=\|Q^\dagger\one_m\|_2$. Hence, each $Z(\xi)$ is a nonempty, compact, and  convex polyhedron.

For $Qz=\xi$ and $b\in\mathcal R$, one has that
$b^\top z=(Q^\dagger b)^\top\xi$. Minimizing first subject
to $Qz=\xi$ and then over $\xi$ yields:
\[
 \begin{aligned}
 \xi^\star(b)
 &=\argmin_{\xi\in\mathcal R}
   \left\{\tfrac12\|\xi-Q^\dagger b\|_2^2+\psi(\xi)\right\}, ~~
 \mathcal S_b=Z(\xi^\star(b)).
 \end{aligned}
\]
In particular, $\mathcal S_b$ is nonempty and compact. Because $r_0$ is the maximum of finitely many linear functions,
the sets $Z(\xi)$ have fixed constraint matrices.
Hoffman's theorem~\cite{hoffman1952} gives a constant $C_H$,
independent of $\xi$, such that
\begin{equation}
 \dist_P(z,Z(\xi))
 \leq C_H\bigl(\|Qz-\xi\|_2
              +\max\{r_0(z)-\psi(\xi),0\}\bigr).
 \label{eq:appendix-fiber-hoffman}
\end{equation}
For any $z\in Z(\xi_1)$, we have $Qz=\xi_1$ and
$r_0(z)=\psi(\xi_1)$. Thus,
\eqref{eq:appendix-fiber-hoffman} gives
$\dist_P(z,Z(\xi_2))
 \leq C_H\bigl(\|\xi_1-\xi_2\|_2+\max\{\psi(\xi_1)-\psi(\xi_2),0\}\bigr)\leq C_H(1+L_\psi)\|\xi_1-\xi_2\|_2$,
where the last inequality uses Lipschitz continuity of $\psi$.
Taking the supremum over $z\in Z(\xi_1)$ bounds the directed
distance from $Z(\xi_1)$ to $Z(\xi_2)$. Interchanging
$\xi_1$ and $\xi_2$ gives the same bound in the reverse
direction. By the definition of Hausdorff distance, $d_H^P(Z(\xi_1),Z(\xi_2))
 \leq C_H(1+L_\psi)\|\xi_1-\xi_2\|_2$.
Non-expansiveness of the proximal map defining $\xi^\star$
then proves~\eqref{eq:analysis-set-lipschitz}, with
$\ell_S=C_H(1+L_\psi)\|Q^\dagger K^{-1/2}\|_2$.

\emph{Uniform residual error bound on bounded regions.}
Set $\xi=Qz$ and $\delta=r_0(z)-\psi(\xi)\geq0$.
Strong convexity gives
\[
 \begin{aligned}
 F_b(z)-\min F_b
 &\geq\tfrac12\|\xi-\xi^\star(b)\|_2^2+\delta,\\
 \dist_P(z,\mathcal S_b)
 &\leq C_H\bigl((1+L_\psi)\|\xi-\xi^\star(b)\|_2+\delta\bigr).
 \end{aligned}
\]
On a bounded region, $0\leq\delta\leq r_0(z)\leq B_r$
for some finite $B_r$, so $\delta^2\leq B_r\delta$.
Consequently, there exists $\mu>0$, independent of $b$,
such that
\begin{equation}
 F_b(z)-\min F_b
 \geq\frac{\mu}{2}\dist_P^2(z,\mathcal S_b).
 \label{eq:appendix-quadratic-growth}
\end{equation}

Let $u=\sfT(z;x)$ and $R=z-u$. For bounded $z$ and
$b$ in the prescribed compact set, $u$ remains bounded;
choose $\mu$ for this image region.
Optimality of the proximal map gives
$(P-\Gamma)R\in\partial F_b(u)$, $|(P-\Gamma)R\|_{P^{-1}}\leq q_K\|R\|_P\leq\|R\|_P$. Next, convexity and~\eqref{eq:appendix-quadratic-growth} imply $\frac{\mu}{2}\dist_P^2(u,\mathcal S_b)
 \leq\|R\|_P\dist_P(u,\mathcal S_b)$.
Using the triangle inequality one gets
\eqref{eq:analysis-residual-error-bound}
with $\kappa=1+2/\mu$.

\emph{Boundedness of the filter trajectory.}
Optimality conditions of the proximal map and positive homogeneity of $r_0$ yield
$\langle z,R\rangle_P
 =F_b(u)+\tfrac12\|Qz\|_2^2
       +\|R\|_P^2-\tfrac12\|QR\|_2^2$. Since $b\in\mathcal R$,
$F_b(u)\geq r_0(u)-\tfrac12\|Q^\dagger b\|_2^2$.
Using global Lipschitz continuity of $r_0$,
$\|QR\|_2^2\leq L\|R\|_P^2$, and $L<2$, completing
the square gives $\langle z,R\rangle_P
 \geq\tfrac12\|Qz\|_2^2+r_0(z)-C_0$,
where $C_0<\infty$ is uniform over the prescribed compact
translation set and independent of $z$ and $\epsilon$.

Along $\epsilon\dot z=-R+e_m$, the component-wise bound
$|e_{m,i}|\leq k_i\overline\eta$ and
\eqref{eq:appendix-fiber-bound} imply, almost everywhere,
\[
 \begin{aligned}
 \frac{\epsilon}{2}\frac{d}{dt}\|z\|_P^2
 &\leq-\tfrac12\|Qz\|_2^2
       +\overline\eta a_1\|Qz\|_2\\
 &\quad-(1-2\overline\eta/\Delta)r_0(z)+C_0.
 \end{aligned}
\]
Because $\overline\eta<\Delta/2$, the right-hand side
tends to $-\infty$ as $\|z\|_P\to\infty$, by
\eqref{eq:appendix-fiber-bound}. A sufficiently large
$P$-ball containing the initial set is therefore forward
invariant.

\subsection{Proof of Proposition~\ref{prop:filter-psd-tracking}}

Since $\underline\beta$ is affine and $x,d$ are absolutely
continuous, the map
$t\mapsto\underline\beta(x(t),d(t))$ is absolutely continuous.
By~\eqref{eq:analysis-set-lipschitz}, for all $s,t\in[0,T]$, $|E_S(t)-E_S(s)|
\leq\|z(t)-z(s)\|_P +\ell_S\|\underline\beta(x(t),d(t))
              -\underline\beta(x(s),d(s))\|_K$.
Consequently, $E_S$ is absolutely continuous.

At almost every $t$ where $E_S(t) > 0$, let $s_t$ be the
projection of $z(t)$ onto $\mathcal S(x(t),d(t))$ based on the norm $\|\cdot\|_P$.
To bound $\dot E_S(t)$, first compare $z(t+h)$ with $s_t$, and then account for the variation of the
set using its Hausdorff distance. Using~\eqref{eq:analysis-set-lipschitz}
one obtains
\[
 \dot E_S(t)
 \leq
 \frac{\langle z(t)-s_t,\dot z(t)\rangle_P}{E_S(t)}
 +\ell_S\nu_\beta(t).
\]

Suppress the time arguments for notational brevity, and write
$R:=z-\sfT(z;x,d)$. By~\eqref{eq:analysis-map-lipschitz} and the condition
$L<2$, $\sfT(\cdot;x,d)$ is non-expansive in $\|\cdot\|_P$. Since
$\sfT(s_t;x,d)=s_t$, we obtain
\[
 \begin{aligned}
 2\langle z-s_t,\sfT(z;x,d)-z\rangle_P
 &=\|\sfT(z;x,d)-s_t\|_P^2\\
 &\quad-\|z-s_t\|_P^2-\|R\|_P^2\\
 &\leq-\|R\|_P^2.
 \end{aligned}
\]
Substituting $\epsilon\dot z=-R+e_m$ into the previous bound on the time derivative, and using the Cauchy-Schwarz inequality and
\eqref{eq:analysis-residual-error-bound}, yields
\[
 \begin{aligned}
 \dot E_S
 \! \leq \! -\frac{\|R\|_P^2}{2\epsilon E_S}
       +\ell_S\nu_\beta+\frac{\|e_m\|_P}{\epsilon}& \! \leq \! -\frac{E_S}{2\epsilon\kappa^2}
       +\ell_S\nu_\beta+\frac{\|e_m\|_P}{\epsilon}\\
 & \! = \! -\frac{\sigma}{\epsilon}E_S
       +\ell_S\nu_\beta+\frac{\|e_m\|_P}{\epsilon}.
 \end{aligned}
\]
On $\{t:E_S(t)=0\}$, absolute continuity and nonnegativity
imply $\dot E_S(t)=0$ almost everywhere, so the same
inequality holds there. Using Gr\"{o}nwall's inequality one can obtain
\eqref{eq:analysis-psd-tracking}.

\subsection{Proof of Proposition~\ref{prop:analysis-closed-loop-tracking}}

Define $r_g(t):=
 G z(t)-G z^\star(x(t),d(t))$.
By~\eqref{eq:analysis-effective-action-tracking},
$\|r_g(t)\|_2\leq gE_{\rm f}(t)$, and
$\dot x= A x + B d + G z^* +r_g$.
Since $\underline\beta$ is affine in $(x,d)$,
almost everywhere,
\begin{equation}
 \begin{aligned}
 \frac{d}{dt}\underline\beta(x(t),d(t))
 &=-C(aI+A)\dot x(t)-C B\dot d(t),\\
 \nu_\beta(t)
 &\leq L_{\beta,x}gE_{\rm f}(t)+v_0.
 \end{aligned}
 \label{eq:analysis-parameter-derivative}
\end{equation}
From Proposition~\ref{prop:filter-pd-tracking} and
Proposition~\ref{prop:filter-psd-tracking}, one can get the bound 
\[
 \begin{aligned}
 \dot E_{\rm f}
 &\leq-\frac{\alpha}{\epsilon}E_{\rm f}
       +\ell\nu_\beta+\frac{\overline e_m}{\epsilon} \leq-\lambda_\epsilon E_{\rm f}
       +\ell v_0+\frac{\overline e_m}{\epsilon}
 \end{aligned}
\]
almost everywhere. Gronwall's inequality yields
\begin{equation}
 E_{\rm f}(t)
 \leq e^{-\lambda_\epsilon t}E_{\rm f}(0)
       +E_\infty(1-e^{-\lambda_\epsilon t})
 \label{eq:analysis-closed-filter-bound}
\end{equation}
where $E_\infty
 :=\frac{\epsilon\ell v_0+\overline e_m}
             {\alpha-\epsilon\ell L_{\beta,x}g}$ and $\lambda_\epsilon:=\alpha/\epsilon-\ell L_{\beta,x}g>0$. Next, note that $\|G z^\star(x,d)
       -G z^\star(y,d)\|_2 \leq
 g\ell L_{\beta,x}\|x-y\|_2$.

Using the definition of $\mathcal T_{\rm act}$, writing the state-error equation in integral form and
using $\|e^{At}\|_2\leq M_Ae^{-\gamma_A t}$ gives
\[
 \begin{aligned}
 E_x(t)\leq{}&
 M_A \e^{-\gamma_A t}E_x(0)\\
 &+\int_0^t \e^{-\gamma_A(t-s)}
 \Bigl[
   \zeta\mathbf{1}_{\mathcal T_{\rm act}}(s)\delta(s)
   +M_AgE_{\rm f}(s)
 \Bigr]\,ds.
 \end{aligned}
\]
Applying the integral form of Gronwall's inequality
to $\e^{\gamma_A t}E_x(t)$ yields
\[
 \begin{aligned}
 E_x(t)\leq{}&
 M_A \e^{-\gamma_A t+\zeta T_{\rm act}(t)}E_x(0)\\
 &+M_Ag\int_0^t
 \e^{-\gamma_A(t-s)
       +\zeta[T_{\rm act}(t)-T_{\rm act}(s)]}
 E_{\rm f}(s)\,ds.
 \end{aligned}
\]
Since $T_{\rm act}$ is non-decreasing and
$T_{\rm act}(s)\geq0$, the last integral is at most $\e^{\zeta T_{\rm act}(t)}
 \int_0^t \e^{-\gamma_A(t-s)}
 \bigl[\e^{-\lambda_\epsilon s}E_{\rm f}(0)
       +E_\infty\bigr]\,ds$.
Finally, $
 \int_0^t e^{-\gamma_A(t-s)}
              e^{-\lambda_\epsilon s}\,ds \leq\frac{1-e^{-\lambda_\epsilon t}}
                {\lambda_\epsilon}$, and 
 $\int_0^t e^{-\gamma_A(t-s)}\,ds
 =\frac{1-e^{-\gamma_A t}}{\gamma_A}$.
Substituting these bounds, dropping  the exponential terms, and using the expressions for
$\lambda_\epsilon$ and $E_\infty$
proves~\eqref{eq:analysis-plant-state-bound}.

\section*{Acknowledgments}

The authors thank Christos Cassandras for the insightful discussions and feedback.

\bibliographystyle{IEEEtran}
\bibliography{biblio}

\end{document}